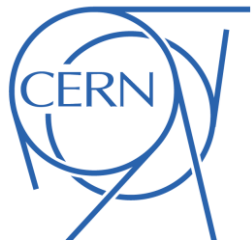 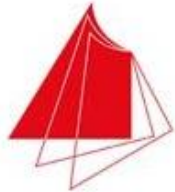

# Bachelor Thesis

## Enhancing Invenio Digital Library With An External Relevance Ranking Engine

**Summer Semester 2012**

Student:              Patrick Oliver GLAUNER
CERN supervisor:      Dr. Tibor SIMKO
Supervising professor: Prof. Holger VOGELSANG
Date:                 2012-06-26

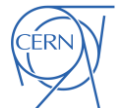

Except where reference is made in the text of this thesis, this thesis contains no material published elsewhere or extracted in whole or in part from a thesis presented by me for another degree or diploma. No other person's work has been used without due acknowledgement in the main text of the thesis. This thesis has not been submitted for the award of any other degree or diploma in any other tertiary institution.

Place and date:

Signature:

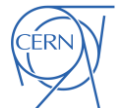

**Acknowledgements**

First and foremost I offer my sincerest gratitude to my supervisor Dr. Tibor Simko who has supported me throughout my thesis with his enthusiasm, patience and expertise. I would also like to thank Prof. Holger Vogelsang for his passion, regular feedback and extremely favorable response time.

I would like to thank the European Organization for Nuclear Research (CERN) for providing an outstanding work environment and a perfectly possible funding.

Furthermore, I am eternally obliged to the members of the Invenio development team for their support and expertise. Every day here is a nice learning experience.

Finally and not to forget, I would like to thank the CERN Restaurant Two staff for preparing coffee in this period. Their coffee kept me writing.

It is a pleasure to stay in the team for the next two years as CERN Fellow.

**Patrick Oliver GLAUNER**

**Title of the Bachelor Thesis**

*Enhancing Invenio Digital Library With An External Relevance Ranking Engine*

**Abstract**

Invenio is a comprehensive web-based free digital library software suite originally developed at CERN. In order to improve its information retrieval and word similarity ranking capabilities, the goal of this thesis is to enhance Invenio by bridging it with modern external information retrieval systems. In the first part a comparison of various information retrieval systems such as Solr and Xapian is made. In the second part a system-independent bridge for word similarity ranking is designed and implemented. Subsequently, Solr and Xapian are integrated in Invenio via adapters to the bridge. In the third part scalability tests are performed. Finally, a future outlook is briefly discussed.

**Patrick Oliver GLAUNER**

**Titel der Bachelor-Thesis**

*Verbesserung der digitalen Bibliothek Invenio durch eine externe Relevance-Ranking-Engine*

**Stichworte**

Dokumentenmanagement, Ranking, Suche, Information Retrieval, Solr, Xapian, elasticsearch, Python

**Zusammenfassung**

Invenio ist eine umfangreiche webbasierte freie digitale Bibliothek, welche ursprünglich am CERN entwickelt wurde. Um dessen Information-Retrieval- und Word-Similarity-Ranking-Fähigkeiten zu verbessern, ist das Ziel dieser Thesis, Invenio durch das Verbinden mit modernen externen Information-Retrieval-Systemen zu verbessern. Im ersten Teil wird ein Vergleich verschiedener Information-Retrieval-Systeme wie z.B. Solr und Xapian durchgeführt. Im zweiten Teil wird eine systemunabhängige Bridge für das Word-Similiarity-Ranking entworfen und implementiert. Im Anschluss werden Solr und Xapian in Invenio mittels Adaptern für die Bridge integriert. Im dritten Teil werden Skalierbarkeitstest durchgeführt. Abschließend wird ein knapper Ausblick geboten.

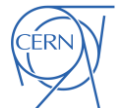

# Contents

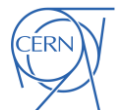

# 1 Introduction

*Document management systems* and *digital libraries* are necessary in any large enterprise or organization to store (document) records and to make it available to staff, clients and users. Record types range from simple text documents through photos to videos. There are not only a lot of functional requirements including submission, *search*, *ranking* and protection features; there is also a significant amount of nonfunctional requirements including *performance*, *scalability* and usability.

This project deals with *indexing*, *searching* and *ranking* of the document management system *Invenio*. Please find general information on Invenio in chapter 1.1. Invenio uses a database-independent metadata indexing module to offer powerful search capabilities. Furthermore, it has a ranking module to rank search result. The existing solution was originally designed for the use case of one million records.

Recent use cases of Invenio include more sophisticated ranking capabilities and up to ten million records. The goal of this project is to take advantage of an existing third-party information retrieval system such as Solr or Xapian and to design a generic bridge between Invenio and them for word similarity ranking. The existing ranking solution should be still available. The native Invenio indexer shall be adapted as far as necessary.

## 1.1 Invenio

Invenio [INV] is a comprehensive web-based free digital library software suite. It has been originally developed at *CERN* in 2002 in the domain of particle physics and related areas. It is now being co-developed by an international collaboration including *CERN*, *Cornell*, *DESY*, *Harvard-Smithsonian* and *the Stanford Linear Accelerator Center*.

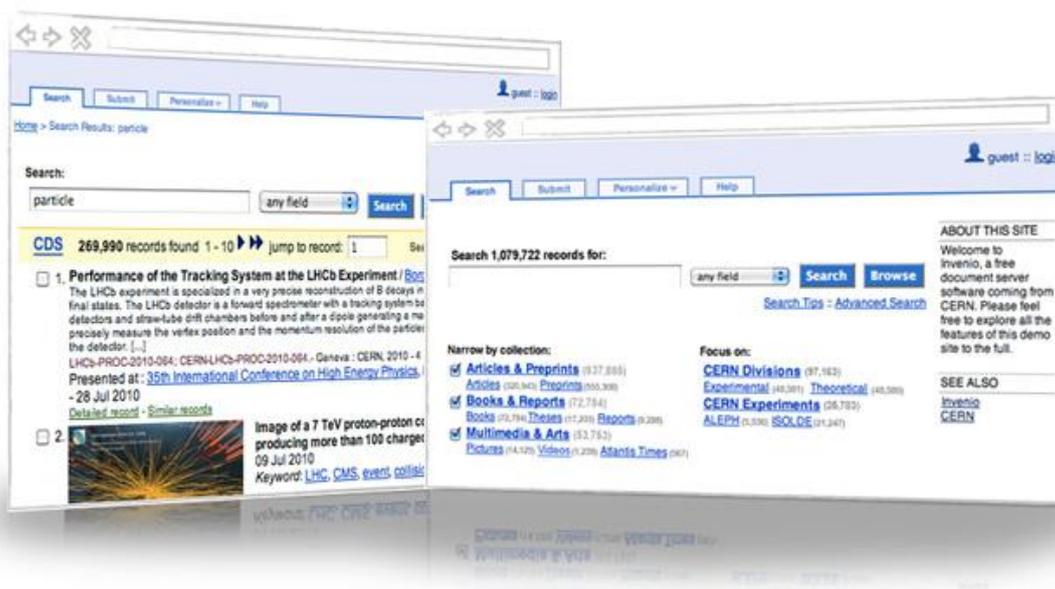

**Figure 1: Sample Invenio screens [INV]**

Invenio is a modular software suite consisting of more than 30 modules. It supports a broad range of features like a navigable collection (i.e. topic) tree, a powerful search engine, flexible metadata, user personalization and standard bibliographic output formats. Furthermore, Invenio supports complex record Ingestion, Curation, Processing and Dissemination processes:

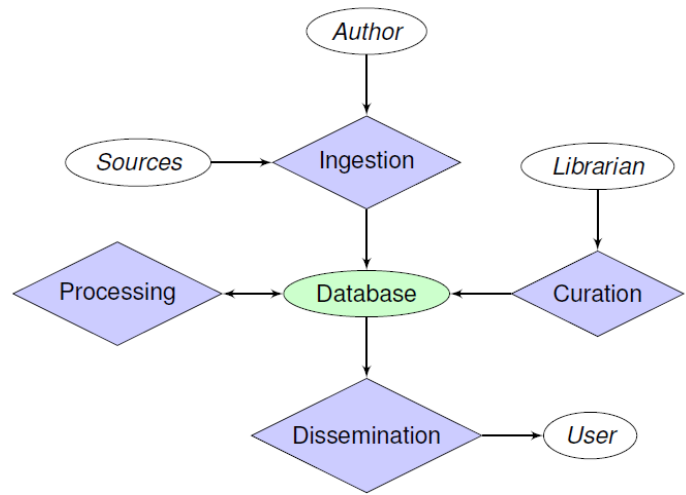

**Figure 2: High-level Invenio overview [SIM]**

There are numerous Invenio instances around the world. The *CERN Document Server* [CDS] is the largest instance with more than one million records:

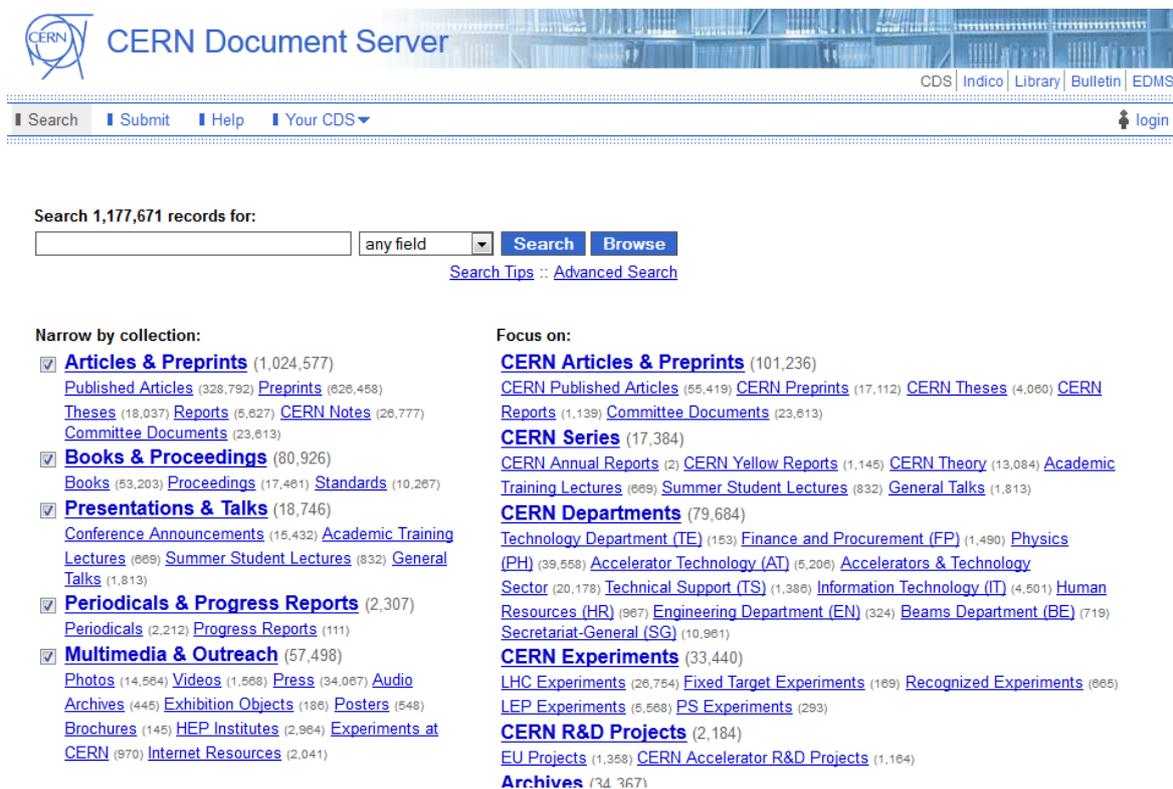

**Figure 3: CERN Document Server [CDS]**

Invenio is primarily written in *Python* with about 300,000 lines of code. The advantage of Python is the high level of abstraction and pseudo code-like programming, not bothering with syntax details or low-level operations. Therefore, Python enables the Invenio developers to fully focus on solving problems within the organic-growth software development model. Some minor parts are written in Java, Javascript, C, XSLT etc. Invenio uses the *MySQL* database system as it complies with the free software policy of Invenio. The Invenio development team uses the source code management system *Git*. Its decentralized development approach meets best the requirements of the world-wide development team. All Invenio modules should have a three-tier architecture i.e. graphical user interface, business logic and database access.

## 1.2 CERN

The *European Organization for Nuclear Research (CERN*) located in Geneva, Switzerland was established in 1954. With more than 10,000 researchers on site, 20 member states and a budget of more than one billion Swiss Francs, CERN is the world's largest particle physics laboratory. The main purpose is to operate the *Large Hadron Collider (LHC)*, the world's largest particle accelerator with a circumference of 27 kilometers.

Driven by the pursuit to push the boundaries of particle physics, CERN focuses strongly on information technology. The most important invention in information technology is the *World Wide Web* invented by Tim Berners-Lee in the late 1980s and early 1990s.

Today, grid computing is a research major to analyze all data generated by the LHC. Further research majors include storage technologies and *collaboration services* including *document management* to improve communication and collaboration services for the world-wide particle physics community.

# 2 Existing solution

This chapter presents the concerned existing parts of Invenio to fulfill the project goal i.e. to take advantage of an existing third-party information retrieval system for word similarity ranking.

## 2.1 Indexing

Invenio uses two kinds of indexes:

1. Regular MySQL database system indexes for columns with high number of selects and lower number of updates like keys and time stamps. This thesis does not discuss them.
2. Invenio database system independent indexes for *metadata* offering more flexible configuration and increased performance to match users' and administrators' requirements. Throughout this thesis they are simply called "indexes" or "fields" and are explained in more detail.

The module creating (Invenio) indexes is called `BibIndex`. Metadata fields are title, author, abstract, collection etc. The index data are saved in database tables. All indexes can be configured in the `BibIndex` admin interface:

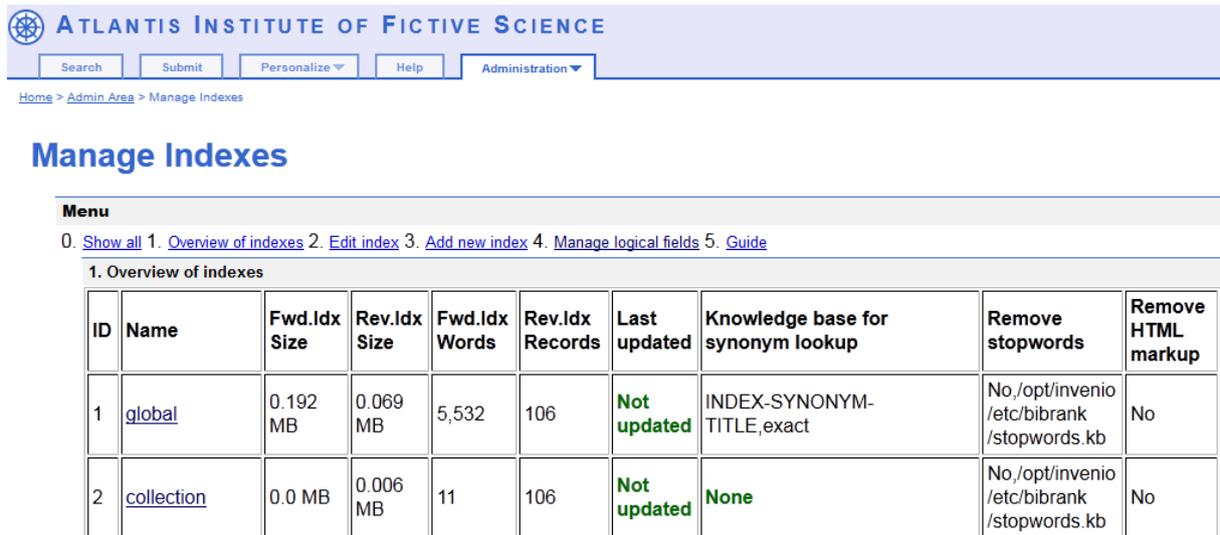

**Figure 4: BibIndex admin interface**

The most significant requirements for indexes include [SIM, 43]:

- Indexes supporting **search** in metadata for words, phrases and regular expressions.
- **Performance** assuming a high amount of selects and a lower amount of updates requiring fast searching and resulting in slower indexation.
- **Index structure** supporting the performance requirements:
  - *Forward indexes* referring from a word to all records containing the word:
    `word1`→`[record4, record5]`
    `word2`→`[record3, record4]`
  - *Reverse indexes* referring from a record to all words contained:
    `record3`→`[word2]`,
    `record4`→`[word1, word2]`,
    `record5`→`[word1]`

- Handling **word frequency** since some words like "the" or "a" have a high frequency whereas most words like "Higgs" are significantly less frequent.

Word frequency is a good example for the necessity of **per-index configuration**. Some metadata like high energy physics abstracts have a high frequency of words like "Higgs" or "Boson" which have nearly no meaning in this case and very little information and should therefore not be indexed. These words are called *stopwords* in Invenio. In contrast, in author metadata a different set of stopwords is necessary. Queries for "Dr. Higgs" should be treated accordingly; therefore "Higgs" should not be a stopword in the author index.

`BibIndex` is executed as a task within the Invenio task scheduler `BibSched`. The execution intervals are set by Invenio administrators depending on specific requirements bearing the long indexing duration in mind.

Most `BibIndex` features, the respective indexing data structures and algorithms are not relevant for this part of the project. Indexing will only be adapted as far as necessary for the use of third-party information retrieval systems for word similarity ranking.

## 2.2 Searching

Invenio offers a powerful search engine implemented in the module `WebSearch`. `WebSearch` uses index data, fulltext search and search in so called "collection clusters". A complex query can easily be built by users:

**Figure 5: Advanced WebSearch search interface**

The following abstract figure shows how a query for the terms "ellis" and "muon" in the collection "thesis" is processed. The index data created by `BibIndex` is used by the fast word search algorithm:

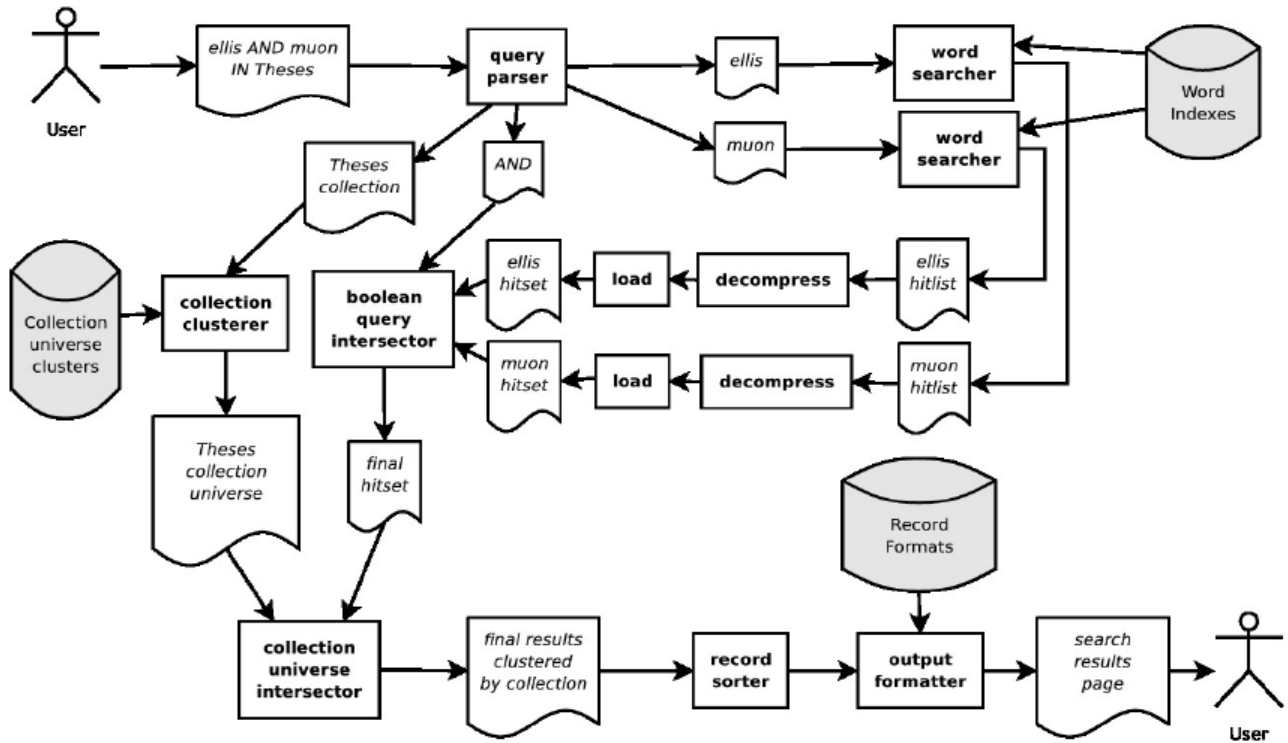

**Figure 6: High-level search processing [SIM, 44]**

The respective search algorithm is not relevant for this part of the project. The returned search results are significantly more important since these have to be ranked.

## 2.3 Ranking

From the usability perspective, users want record search results to be ranked indicating relevance to work more efficiently. The Invenio ranking module is called `BibRank`.

### 2.3.1 Available ranking methods

Invenio currently offers the following ranking methods for records:

- *Sorting* by latest first, title, author, report number and year in either ascending or descending order
- *Ranking* by **word similarity** or *times cited*

Users can select the ranking method in the search interface:

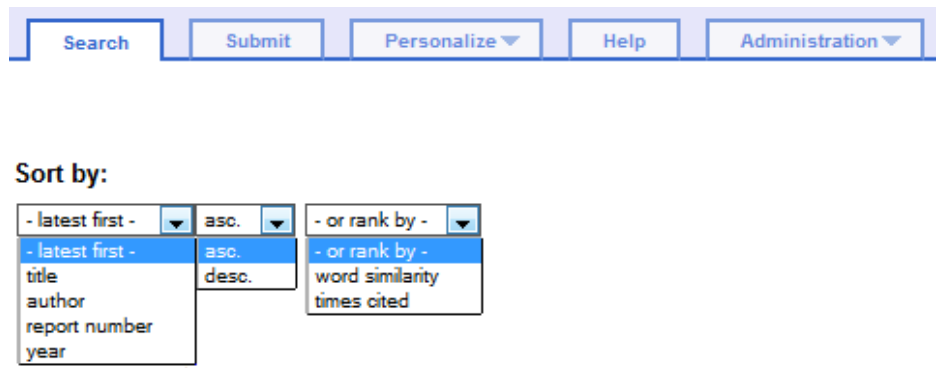

**Figure 7: Search result ranking methods**

Subsequently Invenio displays the ranked results:

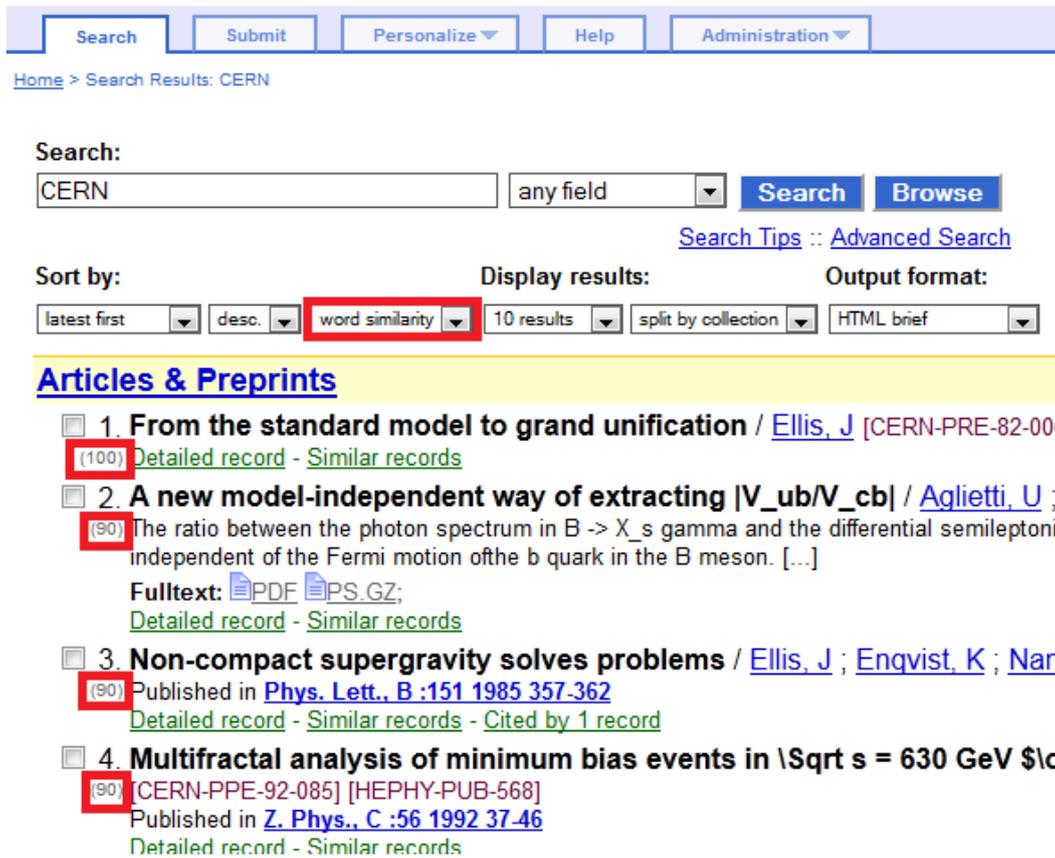

**Figure 8: Ranked search results with percentage display**

`BibRank` is executed as a task within `BibSched` to pre-calculate as much data as possible. These data are saved in the database. The execution intervals are set by Invenio administrators depending on specific requirements bearing the long ranking duration in mind.

### 2.3.2 Word similarity ranking

Since the goal of this project is to take advantage of an existing third-party information retrieval system for **word similarity ranking**, this ranking method is explained briefly but not in depth. The *vector space model* [BAY, 27-30] [CHI] is an algorithm to compute word similarity rankings.

The steps of word similarity ranking are explained in detail in the Invenio `BibRank` word similarity documentation [IWS]. To simplify the computations and to improve ranking quality in Invenio, different preliminary steps are executed: Initially, *stopwords* are removed. Also, *stemming* is used a lot: stemming drops the end of a term reducing the term to its root, e.g. the stem of "reading" is "read" in English.

Subsequently, `BibRank` creates forward and reverse indexes like `BibIndex` (see chapter 2.1) including *weights* indicating "how important the terms are, based on how many times they have been used and how important one term is in one record" [IWS] that are used for later ranking queries. From a simplified point of view, the ranking index structure is:

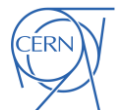

- *Forward indexes*:
  ```
  term1→[(record4,w_f1,4), (record5,w_f1,5)],
  term2→[(record3,w_f2,3), (record4,w_f2,4)]
  ```
- *Reverse indexes*:
  ```
  record3→[(term2,w_r3,2)],
  record4→[(term1,w_r4,1), (term2,w_r4,2)],
  record5→[(term1,w_r5,1)]
  ```

A greater weight means a greater importance of the records or term. There are various methods to calculate the weights depending on word frequency, but those are not necessary from this abstract point of view. In the following examples all weights are **artificial values.** Term weights are explained in detail in [CHI] and [BAY, 27-30].

The *vector space model* represents records and queries as vectors of tuples containing terms and weights. It is explained now in more detail on the basis of an example. Two sample records were indexed and the reverse indexes are:

| Term | Weight |
|------|--------|
| Higgs | 0.3 |
| CERN | 0.2 |
| Science | 0.1 |
| Europe | 0.02 |
| ... | ... |

Table 1: Sample record 1

| Term | Weight |
|------|--------|
| LHC | 0.4 |
| CERN | 0.2 |
| Ranking | 0.1 |
| Higgs | 0.2 |
| ... | ... |

Table 2: Sample record 2

The indexed sample query is:

| Term | Weight |
|------|--------|
| Compiler | 0.1 |
| CERN | 0.4 |
| CMS | 0.5 |
| LHC | 0.2 |
| Higgs | 0.2 |
| Europe | 0.2 |
| ... | ... |

Table 3: Sample query

The vector space model calculates the angle between record and query vectors for the dimensions (terms) that exist in both vectors [CHI, 6]. This has a direct consequence on ranking: the less the angle, the greater the word similarity. Usually, it is more convenient to simply calculate the cosine of the angle between the vectors:

$$cos(\alpha) = \frac{record * query}{\|record * query\|}$$

For record 1, the common terms are "Higgs", "CERN" and "Europe":

$$cos(\alpha_1) = \frac{record_1 * query}{\|record_1 * query\|} = \frac{\begin{pmatrix} 0.3 \\ 0.2 \\ 0.02 \end{pmatrix} * \begin{pmatrix} 0.2 \\ 0.4 \\ 0.2 \end{pmatrix}}{\left\| \begin{pmatrix} 0.3 \\ 0.2 \\ 0.02 \end{pmatrix} \right\| * \left\| \begin{pmatrix} 0.2 \\ 0.4 \\ 0.2 \end{pmatrix} \right\|} = \frac{0.06 + 0.08 + 0.004}{0.36 * 0.49} = 0.82$$

For record 2, the common terms are "Higgs", "CERN" and "LHC":

$$cos(\alpha_2) = \frac{record_2 * query}{\|record_2 * query\|} = \frac{\begin{pmatrix} 0.3 \\ 0.2 \\ 0.4 \end{pmatrix} * \begin{pmatrix} 0.2 \\ 0.4 \\ 0.2 \end{pmatrix}}{\left\| \begin{pmatrix} 0.3 \\ 0.2 \\ 0.4 \end{pmatrix} \right\| * \left\| \begin{pmatrix} 0.2 \\ 0.4 \\ 0.2 \end{pmatrix} \right\|} = \frac{0.06 + 0.08 + 0.08}{0.54 * 0.49} = 0.83$$

Therefore, the ranking is:

1. Record 2
2. Record 1

Invenio uses an advanced variation of the vector space model called *log-entropy weighting scheme* [IWS] [CHI] using also the forward indexes. Additionally, it uses sophisticated weight formulas. This schema is not discussed in this thesis due to lack of space and necessity.

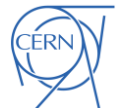

# 3 Requirements and analysis

This chapter presents the requirements and project goal in more detail. Project milestones are identified. Furthermore, there is a presentation and analysis of available third-party information retrieval systems.

## 3.1 Requirements

The existing Invenio word similarity relevance ranking uses an advanced variation of the vector space model. It has several disadvantages:

- The current relevance information is being calculated from the record metadata information only (author, title, abstract etc.).
- The current system is written in Python and hence it is not very fast.
- In Future, it is desirable to extend the current system with more advanced features offered by Natural Language Processing.
- Recent Invenio use cases include up to ten million records and more.

These points shall be partially addressed by taking advantage of an **existing third-party information retrieval system** for **word similarity ranking** permitting to implement enhanced ranking techniques and to improve scalability. A preliminary work has been done within the Invenio project with respect to using the external information retrieval system *Solr* for the use case of fulltext searching. The goal of this project is to cover a generic use case of metadata indexing, **concentrating on the word relevance ranking scenario**.

A bridge between Invenio and third-party external ranking engine shall be created. Invenio can then dispatch word similarity relevance ranking task to the external ranking engine, retrieve its results, and present them to the end user. The designed bridge component shall be **highly generic** so that Invenio installations can easily plug in any of the suitable third-party ranking engines.

## 3.2 Project milestones

The following project milestones have been identified:

- **Study of available third-party information retrieval engines with advanced word similarity relevance ranking capabilities** such as Solr, elasticsearch and Xapian. Compare APIs, performance and scalability.
- **Redesign of word similarity relevance ranking component in Invenio to allow modular plugging of external engines**. Creation of a generic Invenio <-> external engine bridge and implementation of adapter(s) for the most suitable candidate(s) selected in the previous step. Adaption of Invenio native indexer, ranker, and displayer components.
- **Enhancement of ranking configuration of Invenio with a possibility to select any given external ranking engine for any given relevance ranking method**. For example, citation ranking being provided by native Invenio, while word similarity ranking provided by Xapian. Enhancement of indexing configuration as necessary for the word relevance ranking.
- **Performance of scalability measurements** of selected external ranking engines from an index-time perspective (e.g. number of records processed per minute, final index size) and from the search-time end-user perspective (e.g. how fast the retrieved results are ranked).

## 3.3 Available third-party information retrieval systems

*Solr*, *elasticsearch* and *Xapian* are popular open source third-party information retrieval systems. Therefore, they have been identified for possible use in Invenio in this project. In this chapter their main functionality for word similarity ranking is presented and compared including scalability.

### 3.3.1 Apache Solr

Written in Java, *Apache Solr* is built on top of *Apache Lucene* [LUC]. Lucene is written in Java and offers core functionality for data indexing and search. Core Lucene functionality is not discussed in this thesis, only all features offered directly by Solr. Solr is run in a Java application server. All commands like query, add, remove and indexing are sent through HTTP requests and all results are received through HTTP responses. The key strengths of Solr are *advanced search*, *caching* and a *vast variety of different data sources*. Solr offers powerful search capabilities like complex ranking options, spelling suggestions and numeric field statistics. It also supports multi-faceted searching to group results within multiple ranges which can be used in Business Intelligence and many other use cases. Solr offers a broad set of caching techniques including cache warming to preload results of previously executed queries that are likely to be re-executed. Also, caching and warming can be configured by users to improve cache hits and subsequently performance. It offers data import from HTTP, XML, JSON, database systems, Word, PDF etc. This project uses Solr 3.5 [SOL].

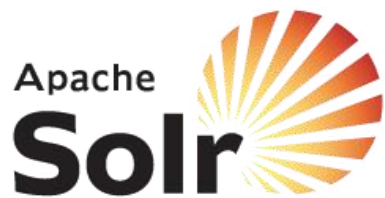

Figure 9: Apache Solr [SOL]

Despite its monumental functionality and amount of features, Solr is *easy to use* and *well-documented*. There is a tutorial [STU] to make first steps. Additionally, there are lots of resources on the Solr webpage and on the web.

Solr is offered as a ZIP file. After downloaded and unzipped, the Jetty application server can be launched with the Solr WAR file via `start.jar`:

```
user:~/apache-solr/example$ java -jar start.jar
```

Finally, Solr is available at `http://localhost:8983/solr/` with an admin interface at `http://localhost:8983/solr/admin/`. The admin interface offers query input fields and various configuration options of the Solr instance:

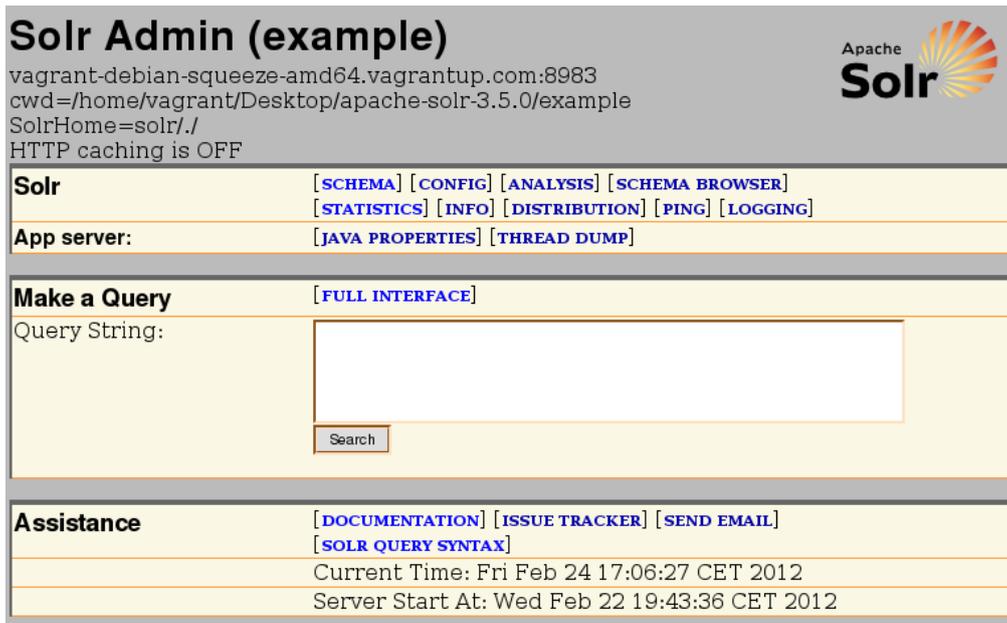

**Figure 10: Solr admin interface**

In general, any Solr statement can be submitted via `post.jar`:

```
user:~/apache-solr/example/exampledocs$ java -jar post.jar STATEMENT
```

Internally, Solr uses a NoSQL-like document store database system. The structure of documents is defined in `schema.xml` which is highly flexible. Solr offers multiple so-called *cores* having separate schema definitions. Therefore, cores allow users to store totally independent types of documents.

```
<fields>
  <field name="id" type="string" indexed="true" stored="true" […] />
  <field title="name" type="text_general" indexed="true" stored="true"/>
  <field name="description" type="text_general" indexed="true" […] />
  […]
</fields>
```

**Source code 1: Solr schema.xml document definition**

Field types are also defined in `schema.xml` which is also highly flexible. For example, different stemmers, stopwords and synonym knowledge bases can be defined for indexing and querying. The following type schema definition is trimmed, also allowing case-sensitivity and further options:

```
<fieldType name="text_general" class="solr.TextField" […] >
  <analyzer type="index">
    <tokenizer class="solr.StandardTokenizerFactory"/>
    <filter class="solr.StopFilterFactory" words="stopwords1.txt" […] />
    <filter class="solr.SnowballPorterFilterFactory" language="English" />
  </analyzer>
  <analyzer type="query">
    <tokenizer class="solr.StandardTokenizerFactory"/>
    <filter class="solr.StopFilterFactory" words="stopwords2.txt" […] />
    <filter class="solr.SynonymFilterFactory" synonyms="synonyms.txt" […]/>
    <filter class="solr.SnowballPorterFilterFactory" language="English" />
  </analyzer>
</fieldType>
```

**Source code 2: Solr field type definition**

In addition, documents added to Solr can have new fields not defined in the schema. Those fields must have pattern definitions called *dynamic fields* defining name including regular expressions and the corresponding field type assignation:

```
<dynamicField name="ignored_*" type="text_general" […] />
<dynamicField name="attr_*" type="text_general" […] />
```

**Source code 3: Solr dynamic field type definition**

Documents are defined in XML and must have a unique key. Documents are added to Solr and are indexed via `post.jar`. The unique key simplifies re-indexing of modified documents.

```
<doc>
   <field name="id">20455-book</field>
   <field name="title">Core Java</field>
   <field name="description">A must have for Java programmers</field>
   […]
</doc>
```

**Source code 4: Sample Solr document**

```
user:~/apache-solr/example/exampledocs$ java -jar post.jar document.xml
```

By default, Solr HTTP query responses are XML with other response types like JSON possible. XML and JSON are non-convenient formats for Python programming since Python types like dictionaries are more preferable. *solrpy* [SPY] is a well-documented and easy to use Python wrapper for Solr making it an ideal candidate for the project. A sample source code for adding and querying data demonstrates this:

```python
import solr
# create a connection to the solr server
s = solr.SolrConnection('http://localhost:8983/solr')
# add a document
s.add(id='20455-book', title='Core Java', description='A must […] ')
s.commit()
# query all documents with title Java, search in field
response = s.query('title:Java')
for hit in response.results:
        print hit['title']
```

**Source code 5: solrpy sample code**

Solr offers easy to use *word similarity ranking* which uses internally Lucene's vector space model implementation [SSI]. For the previously defined schema, the following sample documents are added. These data are not meant to represent a digital library. This tiny document set is **illustrative** and shall help to comprehend and evaluate the following ranking results. It shall also help to find possible configuration difficulties.

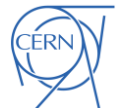

| id | title | description |
|---|---|---|
| 20455-book | Core Java | Best Java book for programmers to learn writing Java programs. |
| 24004-book | Effective Java | Yet another must have for every Java programmer. Written by a Java expert for Java programmers to write better Java programs. |
| 2021-movie | Java Beaches | A movie about beaches of the island of Java. The beaches are white and surrounded by clear water. Not for programmers. |
| 42-podcast | How to become a programmer | Introduction to different programming languages including Python, C++ and ABAP. |

<div align="center">Table 4: Sample data for word similarity ranking</div>

All documents containing the terms "Java" and "Programmer" in their description shall be returned and ranked by word similarity. Using solrpy, this query can easily be written:

```
response = s.query('description:Java AND description:Programmer',
           fields=['score', 'title'], sort="score", sort_order="desc")
```

<div align="center">Source code 6: solrpy word similarity query</div>

The query returns a list of ranked result titles including the internally calculated ranking scores:

| Score | Title |
|---|---|
| 0.533471.. | Core Java |
| 0.452666.. | Effective Java |
| 0.265165.. | Java Beaches |

<div align="center">Table 5: solrpy word similarity result</div>

With *English stemming* and *case-insensitivity* enabled using *no stopwords* in the schema, some queries return exactly the same result and scores, e.g. for "java" and "programmers". The result is analyzed and interpreted in more detail from an implementation-independent view:

The description of **"Core Java"** contains ten terms with two of them being "Java". One of them is "programmers" having the root "programm". Therefore, a query for "Programmer" or "Programmers" returns the same scoring since both of them have the same root considering case-insensitivity. The description also contains the term "programs" having the root "program" and is therefore not taken into account. In conclusion, the query terms have a very high frequency within the description.

The description of **"Effective Java"** contains 21 terms with four of them being "Java". Two of them are "programmers". Therefore, the ranking score is close to the one of "Core Java".

The description of **"Java Beaches"** contains also 21 terms with one of them being "Java" and one being "programmers". Therefore, the ranking score is significantly less than the ones of "Core Java" and "Effective Java".

The description of **"How to become a programmer"** does not contain the term "Java" and is therefore not in the result set.

Using stopwords like "to" and "for" could further improve the ranking scores.

In conclusion, Solr offers advanced and easy to use information retrieval capabilities. It is well-documented and can be configured flexibly. Its word similarity ranking capabilities are easy to use and deliver reasonable results. Solr can easily be used in Python through the solrpy wrapper.

### 3.3.2 elasticsearch

*elasticsearch* is another Java search server built on top of Apache Lucene. Like Solr, it is run in a Java application server. All commands are also sent through HTTP requests. The development team has only one main developer. The key strengths of elasticsearch are a *schema-less data model*, *multi tenancy*, *scalability for distributed computing* and *near real-time search*. Since elasticsearch is schema-less unless explicitly defined, any JSON documents can be added to elasticsearch. This results in a great flexibility and low configuration effort. Elasticsearch offers so-called "multi tenancy" allowing documents to be divided into separate "indexes". An index is equivalent to a database in relational database systems. elasticsearch offers "sharding" allowing indexes to be broken up into smaller "shards". Replication of indexes and shards allows a distributed environment with copies on different nodes. Another design goal of elasticsearch is near real-time search also supported in multi tenant and multi node environments. This project uses elasticsearch 0.18.7 [ELA].

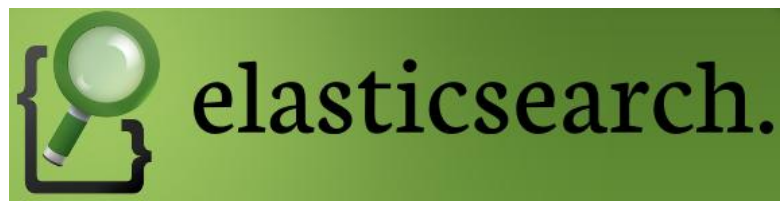

There are different resources on the elasticsearch webpage [ELA] including a guide, tutorials and videos to make first steps. In general, the documentation is insufficient but there are more resources on the web like forums and mailing list archives.

elasticsearch is offered as a ZIP file. After downloaded and unzipped, it can be launched in foreground mode via a script. Finally, Solr is available on port 9200 at `http://localhost:9200/`.

```
user:~/elasticsearch/bin$ ./elasticsearch -f
```

In general, any elasticsearch statement can be submitted via the Linux command line tool `curl`:

```
$ curl -XPUT http://localhost:9200/index/type/id -d STATEMENT
```

Internally, elasticsearch uses a NoSQL database system supporting JSON documents. Due to multi tenancy, an elasticsearch instance can have more than one *index*. An index consists of *types* which consist of *fields*. In accordance to relational database systems, an index is equivalent to a database, a type is comparable to a table and a field is comparable to a column. No type schema definition is necessary which makes elasticsearch highly flexible. Indexes and types are created automatically if they do not exist yet. Field data types are "automatically detected and treated accordingly" [ELA]. Schemas can be defined to configure elasticsearch types to match requirements more precisely. Different indexes can be used for multiple reasons. For example, multiple indexes can be used for documents of different applications like library records, phonebooks and code repositories. There are more use cases possible like different indexes for privacy protection.

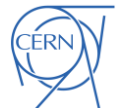

Documents are added to Solr and indexed via `curl` and must have a unique key. The unique key simplifies re-indexing of modified documents. In this example the document "20455-book" is added to the *type* "record" within the *index* "Invenio" with the two *fields* "title" and "description":

```
$ curl -XPUT http://localhost:9200/invenio/record/20455-book -d '{
      title: "Core Java",
      description: "A must have for programmers"
}'
```

<div align="center">Source code 7: Sample elasticsearch document</div>

All index configurations are also sent through HTTP requests. There are various possibilities to configure elasticsearch. For example, *analyzers* consisting of stemmers, tokenizers and further options like synonyms and stopwords can be defined for indexing and searching. Unfortunately, the construction of complex analyzers is poorly documented like most elasticsearch functionality. In the following example the defined analyzers are used per index:

```
$ curl -XPOST localhost:9200/invenio -d '{
   "index" : {
      "analysis" : {
         "index analyzer" : {
            "tokenizer" : "standard",
            "filter" : ["standard", "lowercase", "porter_stem"]
         },
         "search analyzer": {
            "tokenizer" : "standard",
            "filter" : ["standard", "lowercase", "porter_stem"]
}}}}'
```

<div align="center">Source code 8: elasticsearch analyzer definition</div>

Schemas *may* be defined per type within an index to match requirements more precisely. The set of properties for a type is called a *mapping* defining properties like fields, their corresponding data type and analyzers. In the following example the two fields "title" and "description" are defined. They could have separate analyzer settings, but in this example they use the previously defined per-index analyzers.

```
$ curl -XPUT localhost:9200/invenio/record/_mapping -d '{
   "record" : {
      "properties" : {
         "title" : {"type":"string", "store" : "yes"},
         "description" : {"type" : "string", "store" : "yes"}
}}}'
```

<div align="center">Source code 9: elasticsearch mapping definition</div>

*PyES* [PYE] is an easy to use Python wrapper for elasticsearch making it interesting for the project. Unfortunately, it is poorly-documented and has some bugs making it difficult to use. A sample source code for adding and querying data demonstrates its basic functionality. A response is a nested Python dictionary containing documents and further information on ranking and timing which is not relevant for this part of the project.

```python
import pyes
# create a connection to the eslasticsearch server
conn = pyes.ES('localhost:9200')
# add a document
conn.index({ "title":"Core Java",                        # data
             "description":"A must have for programmers"}, # data
             "invenio",                                   # index
             "record",                                    # type
             "20455-book")                                # id
conn.refresh()
# query all data with title Java, search in field
q = pyes.StringQuery("title:Java")
results = conn.search(query = q)
print results
```

<p align="center">Source code 10: pyes sample code</p>

PyES can also be used for index configuration and mapping. Unfortunately, there is no sufficient API documentation available. Therefore and in addition to the lack of the corresponding elasticsearch documentation, this feature was not evaluated.

elasticsearch offers to write *word similarity ranking* queries using Lucene functionality easily, whereas the construction of sufficient analyzers is extremely difficult due to the lack of documentation. For the previously defined schema, the documents defined in chapter 3.3.1 shall be added. All documents containing the terms "Java" and "Programmer**s**" in their descriptions shall be returned and ranked by word similarity using PyES. This query can easily be written:

```
q = StringQuery(query = "description:Programmers AND description:Java")
results = conn.search(query = q)
```

<p align="center">Source code 11: PyES word similarity query</p>

The query returns a dictionary of results including the internally calculated ranking scores:

| Score | title |
|---|---|
| 0.53347... | Core Java |
| 0.31529... | Effective Java |
| 0.21019... | Java Beaches |

<p align="center">Table 6: PyES word similarity result</p>

Sorting the results by their ranking score results in the same order returned from Solr in chapter 3.3.1. Using stopwords like "to" and "for" could even further improve the ranking scores. Lowercase queries return exactly the same result. In accordance to the index-wide analyzer definition, a query for "Programmer" instead of "Programme**rs**" **should** return exactly the same result, which it **does not**. Due to the lack of documentation it was neither possible to realize the correct behavior nor to find the reason for the behavior in the analyzer definition. It seems like the indexed document data is not stemmed as demanded.

In conclusion, elasticsearch offers advanced information retrieval capabilities that are partially easy to use. It is poorly documented due to the small development team. Therefore, most capabilities cannot be fully used and some query results are not reasonable. Second, it only supports JSON and no broad set of input formats. The python wrapper PyES is also poorly documented which makes only its basic functionality usable.

### 3.3.3 Xapian

Written in C++, *Xapian* is another open source information retrieval system. Primarily, it is designed for fulltext search. It is a *probabilistic information retrieval* engine [BAY, 30-34] not using the vector space model. Due to lack of necessity the probabilistic model is not explained in this thesis. It has some advantages but the "vector model is expected to outperform the probabilistic model with general data" [BAY, 34]. The key strengths of Xapian are *scalability*, *performance* and *portability*. Since Xapian is written in C++ it offers high performance for even hundreds of millions records and scales therefore very well. Despite its C++ code base it is available for many platforms including various Linux distributions, Mac OS X and Windows. Furthermore, it offers language bindings for many languages including Java, Python, C# and PHP. This project uses Xapian 1.2.8 [XAP].

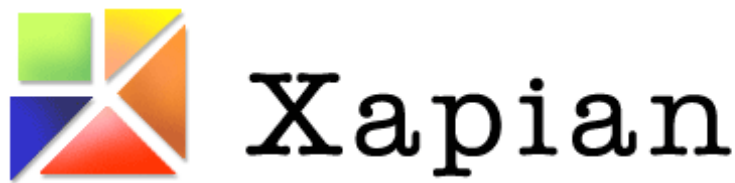

<div align="center">

**Figure 12: Xapian [XAP]**

</div>

Despite its monumental functionality and amount of features, Xapian is *easy to use* and *well-documented*. There are tutorials and further resources on the Xapian webpage and on the web. Xapian and language bindings can be easily installed. Most features are already included in Debian 6. All Xapian functionality is accessible through a C++ library. All code examples in this thesis use directly the Python wrapper `python-xapian` that is available on the Xapian webpage [XAP].

Internally, Xapian can use and manage multiple document store databases in parallel. A Xapian database is a set of documents. Each document consists of one string for performance reasons. There is neither a schema definition nor are multi value documents supported yet [XAT]. At this part of the project, only descriptions are added with a more general discussion at the end of the Xapian analysis. In this example the document "20455" is added to the database "xapiantest". Only unsigned non-zero document ids are supported by Xapian. Stemmers, possible stopwords etc. are defined directly in the Python source code and not in any configuration file or schema.

```python
import xapian
# open the database for updating, creating a new database if necessary
database = xapian.WritableDatabase("xapiantest", xapian.DB_CREATE_OR_OPEN)
indexer = xapian.TermGenerator()
stemmer = xapian.Stem("english")
indexer.set_stemmer(stemmer)

content_string = "A must have for programmers"
doc = xapian.Document()
doc.set_data(content_string)
# index document
indexer.set_document(doc)
indexer.index_text(content_string)
# add/replace document
database.replace_document(20455, doc)
```

**Source code 12: python-xapian sample insert**

Executing the program creates the database "xapiantest" in a subfolder of the script's location. The construction of queries is very similar to the former insertion example with separate configurations including stemmer etc.:

```python
import xapian
# open the database for searching
database = xapian.Database("xapiantest")
# start an enquire session
enquire = xapian.Enquire(database)
# query top 10 results with containing Java
query_string = "Java"
qp = xapian.QueryParser()
stemmer = xapian.Stem("english")
qp.set_stemmer(stemmer)
qp.set_database(database)
qp.set_stemming_strategy(xapian.QueryParser.STEM_SOME)
query = qp.parse_query(query_string)
enquire.set_query(query)
matches = enquire.get_mset(0, 10)

for m in matches:
    print "%i: %i%% docid=%i [%s]" % (m.rank + 1, m.percent, m.docid,
                    m.document.get_data())
```

**Source code 13: python-xapian sample query**

Xapian offers easy to use to use *word similarity ranking* capabilities. In accordance to chapter 3.3.1, the defined documents shall be added. All documents containing the terms "Java" and "Programmers" in their descriptions shall be returned and ranked by word similarity. All terms are stemmed except of those starting with a capital letter (see documentation of `set_stemming_strategy()`). Since Xapian search is case-insensitive by default, the query can easily be written including all configurations:

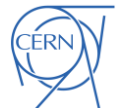

```
query_string = "java AND programmers"
qp = xapian.QueryParser()
stemmer = xapian.Stem("english")
qp.set_stemmer(stemmer)
qp.set_database(database)
qp.set_stemming_strategy(xapian.QueryParser.STEM_SOME)
query = qp.parse_query(query_string)
enquire.set_query(query)
matches = enquire.get_mset(0, 10)
```

**Source code 14: python-xapian word similarity query**

Subsequently, it returns a list of ranked results including the internally calculated ranking scores:

| score | title |
|-------|-------|
| 100% | Core Java |
| 89% | Effective Java |
| 65% | Java Beaches |

**Table 7: python-xapian word similarity result**

With *English stemming* enabled and *case-insensitivity* search by default using, some queries return exactly the same result and scores, e.g. for "java" and "programmer" or "java" and "programm". Invoking the stemmer can help to understand results better: `stemmer(query)` returns the stemmed query.

Xapian documents are single strings primarily supporting efficient fulltext search. Multiple values are not supported yet. This feature was requested more than four years ago. It has not been realized yet due to significant necessary changes within Xapian and API backward compatibility concerns [XAT]. Invenio needs to save multiple metadata values per record like author, title and abstract. There are different approaches to this problem:

First, all record metadata could be saved in a Python dictionary:

```
record = {"title": "...",
          "abstract": "...",
          "author": "..."
}
```

**Source code 15: Record metadata as Python dictionary**

This dictionary can be easily converted into a string and then be completely added as one document. Possible query results can be converted easily into Python dictionaries with `eval(dictionary)`. This approach might result in a performance decrease. Queries would have to make use of complex regular expressions to guarantee correct results. Unfortunately, regular expressions are not supported by Xapian. This could be implemented in Python which would result in an even worse performance decrease.

Second, metadata values could be separated by separators that are guaranteed not to be within the values. This would make the construction of queries inconvenient. Therefore, a wrapper for saving and querying of multiple values should be implemented. This wrapper could offer field search syntax comparable to Solr syntax.

Third, a separate database per metadata field could be created and used in parallel. This might result in a performance decrease and more Python code to calculate intersections of separate result sets.

Finally, Omega [OME] is a search library built on top of Xapian offering support to handle various document types including HTML, PHP and PDF. With some extra effort, it could also be used to work on metadata represented in XML format.

Therefore, the lack of multi value support requires further analyses and additional programming. This topic shall not be discussed further in this part of the project. In case a Xapian adapter is implemented in the later part of the project, this has to be a preliminary task.

In conclusion, Xapian offers advanced and easy to use information retrieval capabilities. It is well-documented and can be configured very flexibly directly in the program code. Xapian configuration in the program code implies an improved testability and transparency of results. The lack of multiple value support makes queries less convenient and enforces workarounds. Its word similarity ranking capabilities are easy to use and deliver reasonable results. Xapian can easily be used in Python through the python-xapian wrapper.

### 3.3.4 Initial feasibility study

Due to the lack of the elasticsearch documentation and the resulting non-reasonable query results, it is not further evaluated. Therefore, Solr and Xapian including their Python wrappers are evaluated in this chapter considering their resource consumption as the document amount increases. The test data are the abstracts of the 1000 most recently added theses on the CERN Document Server [CDS]. With 1000 abstracts, this study **shall only convey an illustrative surface impression of Solr and Xapian** tendency. A detailed study with more data is only useful as soon as possible adapters for Solr and Xapian are created that can subsequently be optimized.

All previously used configurations are used for Solr and Xapian. Since Solr and Xapian are configured differently, they are **only compared with themselves** and not with the respective other. Building an (almost) similar configuration is time-consuming and could easily require a few months or might not be feasible at all. Replicas and distributed environments are not tested due to lack of necessity in this analysis.

The amount of terms (words separated by whitespaces) and characters of the test data is well-distributed:

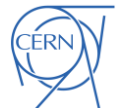

| Abstracts of the most recently added theses | Term count | Character count |
|---|---|---|
| 1-100 | 22,707 | 147,739 |
| 101-200 | 25,859 | 168,282 |
| 201-300 | 27,167 | 173,382 |
| 301-400 | 29,480 | 187,289 |
| 401-500 | 28,566 | 182,782 |
| 501-600 | 30,355 | 198,276 |
| 601-700 | 27,986 | 178,006 |
| 701-800 | 30,048 | 189,480 |
| 801-900 | 28,609 | 183,333 |
| 901-1000 | 25,928 | 166,494 |
| ∑ | 276,705 | 1,775,063 |
| Distinct count | 26,407 | - |

Table 8: Tendency test data term and character counts

In all tendency studies, the following commands are executed for the 100, 200, 300, 400, 500, 600, 700, 800, 900 and 1000 most recently added theses:

1. Insertion of the abstracts into an empty database
2. Updating all abstracts with exactly the same content
3. Querying all abstracts for the conjunction of the terms "CERN", "Higgs" and "Boson"

All commands are executed ten times in row and the average duration is calculated. The test system is a Debian "Squeeze" 64 bit Virtual Box image with 2 CPUs and 1500 MB RAM. Background jobs could distort the measurements since it is not a specially prepared test system. Delete commands are not evaluated since they are rarely executed in Invenio use cases.

**Solr:** The increase from 100 to 1000 abstracts results in an insertion duration that is approximately 10 times longer. The updating durations are nearly the same since Solr always overrides records to be updated. Therefore, insertion and updating durations of Solr develop very well for this amount of data:

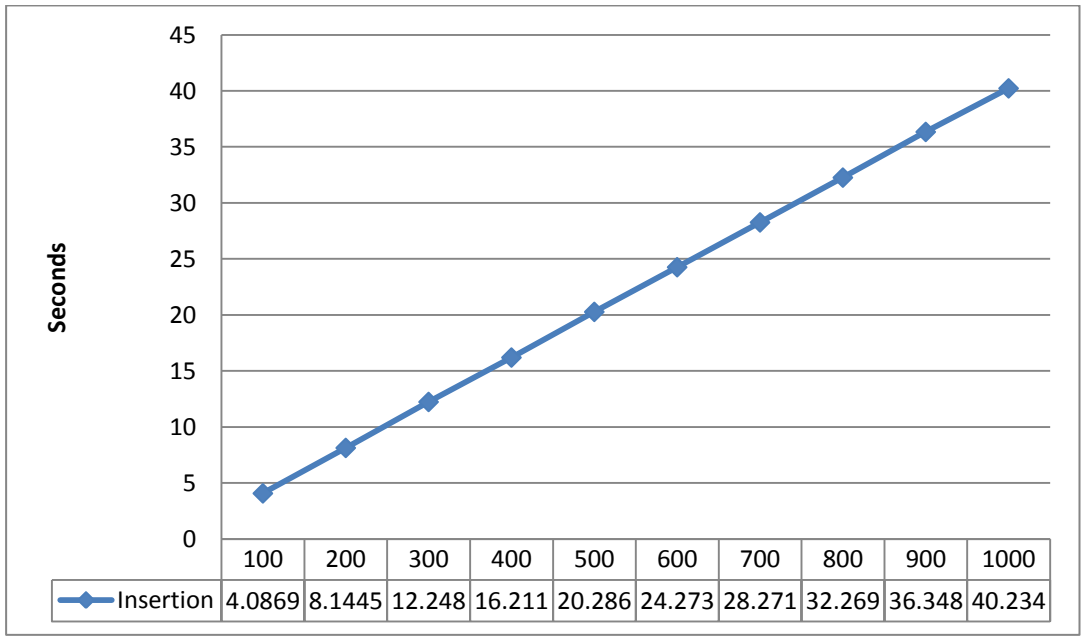

|  | 100 | 200 | 300 | 400 | 500 | 600 | 700 | 800 | 900 | 1000 |
|---|---|---|---|---|---|---|---|---|---|---|
| Insertion | 4.0869 | 8.1445 | 12.248 | 16.211 | 20.286 | 24.273 | 28.271 | 32.269 | 36.348 | 40.234 |

**Figure 13: Solr insertion duration trend**

The index size develops slightly worse since the increase from 100 to 1000 abstracts results in an index size that is approximately 12 times greater:

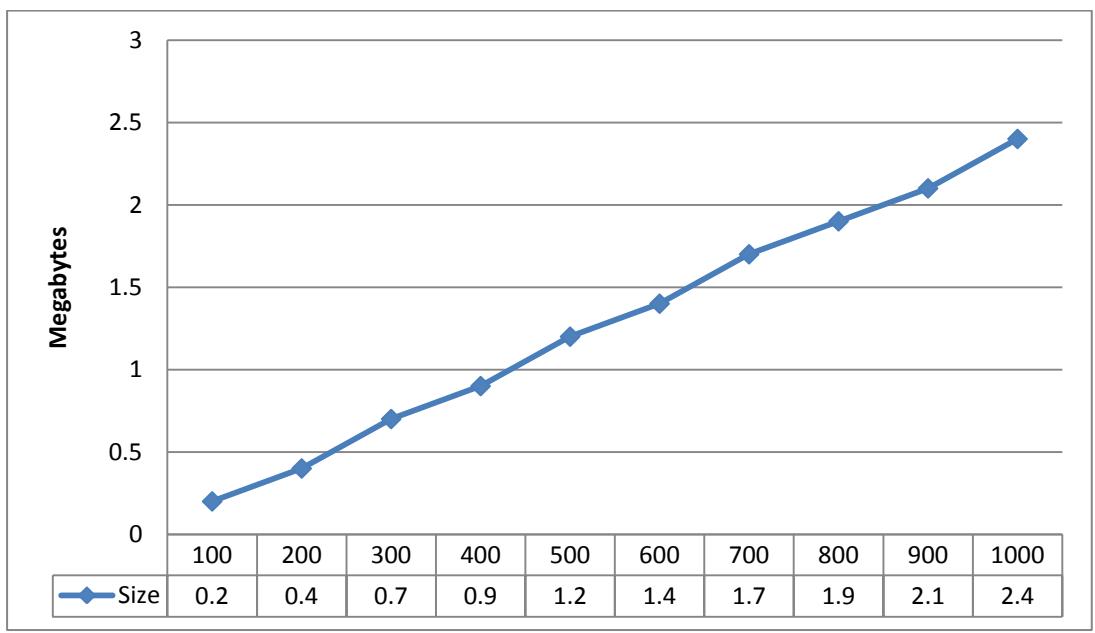

|  | 100 | 200 | 300 | 400 | 500 | 600 | 700 | 800 | 900 | 1000 |
|---|---|---|---|---|---|---|---|---|---|---|
| Size | 0.2 | 0.4 | 0.7 | 0.9 | 1.2 | 1.4 | 1.7 | 1.9 | 2.1 | 2.4 |

**Figure 14: Solr index size trend**

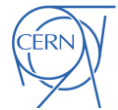

Querying all abstracts for the conjunction of the terms "CERN", "Higgs" and "Boson" results in the following result counts:

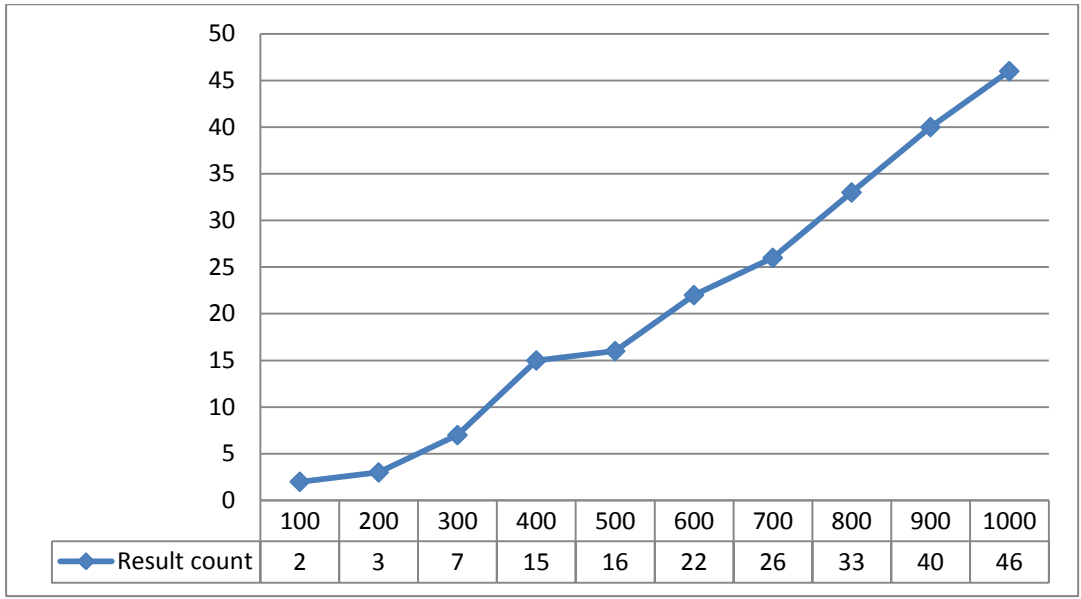

| | 100 | 200 | 300 | 400 | 500 | 600 | 700 | 800 | 900 | 1000 |
|---|---|---|---|---|---|---|---|---|---|---|
| Result count | 2 | 3 | 7 | 15 | 16 | 22 | 26 | 33 | 40 | 46 |

**Figure 15: Solr result count trend**

All word similarity **queries** are performed in about 40ms each. Query tendency studies are not possible for only 1000 abstracts since the results might easily be distorted by background activity.

**Xapian**: The increase from 100 to 1000 abstracts results in an insertion duration that is approximately 14 times longer. It also results in an updating duration that is approximately 12 times longer. Therefore, insertion and updating durations of Xapian develop well for this amount of data:

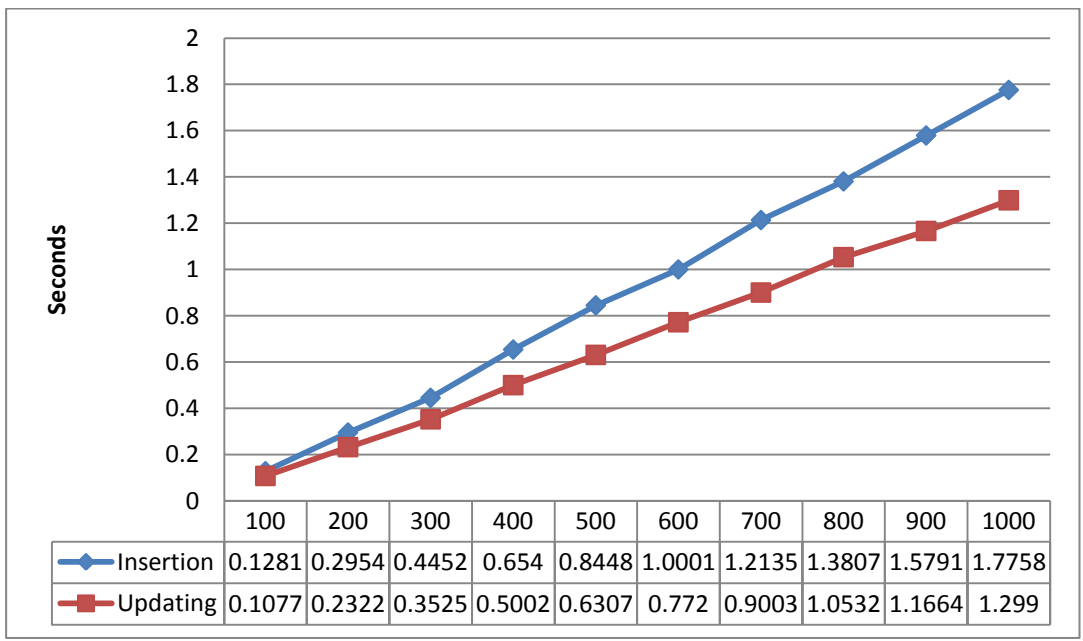

| | 100 | 200 | 300 | 400 | 500 | 600 | 700 | 800 | 900 | 1000 |
|---|---|---|---|---|---|---|---|---|---|---|
| Insertion | 0.1281 | 0.2954 | 0.4452 | 0.654 | 0.8448 | 1.0001 | 1.2135 | 1.3807 | 1.5791 | 1.7758 |
| Updating | 0.1077 | 0.2322 | 0.3525 | 0.5002 | 0.6307 | 0.772 | 0.9003 | 1.0532 | 1.1664 | 1.299 |

**Figure 16: Xapian insertion and updating duration trend**

The index size develops slightly better since the increase from 100 to 1000 abstracts results in an index size that is approximately 9 times greater:

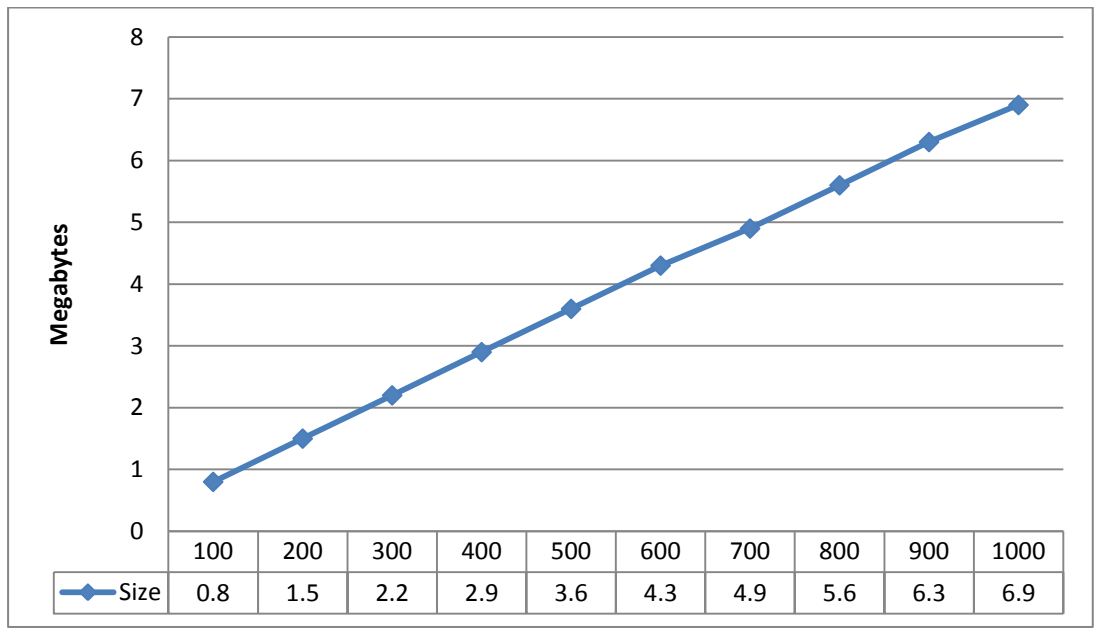

| | 100 | 200 | 300 | 400 | 500 | 600 | 700 | 800 | 900 | 1000 |
|---|---|---|---|---|---|---|---|---|---|---|
| ◆ Size | 0.8 | 1.5 | 2.2 | 2.9 | 3.6 | 4.3 | 4.9 | 5.6 | 6.3 | 6.9 |

**Figure 17: Xapian index size trend**

All word similarity **queries** are performed extremely quickly and take about 0.1ms each. Query tendency studies are not possible for only 1000 abstracts since the results might easily be distorted by background activity. The current Xapian configuration returns exactly the same result sets as Solr.

**elasticsearch** promises to support near real-time search also in multi tenant and multi node environments [ELA]. In environments searching in parallel to index updates, it could even demonstrate to be up to ~50 times faster than Solr [ESV].

In conclusion, Solr develops better for insertion and updating, whereas the Xapian index size develops better. The following table summarizes the factor increases from 100 to 1000 abstracts:

| | Solr | Xapian |
|---|---|---|
| **Insertion duration** | 10 | 14 |
| **Updating duration** | 10 | 12 |
| **Index size** | 12 | 9 |
| **Querying duration** | Not measureable | Not measureable |

**Table 9: Comparison of Solr and Xapian tendency**

Independently, Xapian seems to be significantly faster than Solr. There are multiple possible reasons for this effect. First, Xapian code accesses its local database directly whereas Solr uses HTTP requests and responses which might result in a performance decrease due to application server settings. Second, they are configured differently which might have an impact on performance. Third, both Python wrappers could affect performance e.g. different or no use of parallelization. Fourth, port forwarding in the virtual machine settings might cause a slowdown. Fifth, the parallelization of HTTP requests might increase the performance.

The Xapian index size is greater than the Solr one. Solr keeps 10,211 of the 26,407 distinct terms, whereas Xapian keeps 22,901. Solr makes greater use of stemming and keeps only the word roots. For example, for all abstracts, Solr keeps only the terms "creat" and "creation". In contrast, Xapian keeps the following terms: "create", "created", "creates", "creating", "creation" plus "**Z**creat" and "**Z**creation" with the Z prefix being an internal format for stemming.

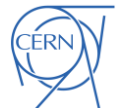

### 3.3.5 Comparison

In this chapter the necessary functionality of the previously presented and analyzed information retrieval systems for word similarity ranking is compared. Implications on the development of information retrieval system adaptors for the future generic bridge are also discussed.

Solr and elasticsearch are written in Java and built on top of Lucene which is also written in Java. Therefore, both of them are highly **platform independent**. Despite its C++ code base, Xapian is available for many different platforms and therefore also highly platform independent.

Solr and Xapian are well-**documented** and hence easy to use. In contrast, elasticsearch is poorly-documented which makes the implementation of real-word use cases unpredictable. The Solr and Xapian **Python** adapters are also well-documented and can be used very well. In accordance to the poor elasticsearch documentation, its Python wrapper is also poorly-documented.

The Solr **data format** is defined in the `schema.xml` file also offering flexibility so that documents can have more elements. A Solr instance can contain more than one core having each a separate schema. elasticsearch offers the most flexible data model. A schema definition is possible but not necessary since elasticsearch accepts any reasonable JSON input. Xapian accepts only one single string as document and no multi value input which limits its capabilities. In addition, there are possibilities and workarounds to support multi value documents that require further efforts for a possible Xapian adapter. Furthermore, Xapian only accepts unsigned non-zero integer ids which is less flexible but sufficient for the current Invenio version supporting only positive integer record ids.

Solr is also **configured** in the `schema.xml` file. Its capabilities are well-documented making changes to the schema very easy. In addition, an admin GUI makes configuration and testing easy. elasticsearch is fully configured through HTTP JSON requests. The exact capabilities and how to use them are not documented in detail making the configuration difficult producing unpredictable results. Xapian is fully configured in the program code. This kind of configuration is convenient and improves testability and allows additional configuration depending on the control flow. Its configuration (capabilities) is well-documented.

Solr and elasticsearch offer vector space model word **similarity ranking capabilities**. In contrast, Xapian is based on a probabilistic model. All three of them are easy to use. Solr and Xapian produce reasonable results. In contrast, the results of elasticsearch are not fully reasonable due to the lack of configuration documentation.

**Resource consumption tendency** factors of Solr and Xapian are compared and illustrated in chapter 3.3.4.

In the following table, the properties of Solr, elasticsearch and Xapian are summarized and ranked:

| | Solr | elasticsearch | Xapian |
|---|---|---|---|
| **Platform independence** | + | + | + |
| **Documentation** | + | - | 0 |
| **Python integration** | + | - | + |
| **Data format definition** | 0 | + | - |
| **Configuration** | 0 | - | + |
| **Word similarity ranking** | + | + | + |

Table 10: Information retrieval systems comparison and ranking

In conclusion, Solr has a very positive overall evaluation and is already partly integrated into Invenio. Therefore, the implementation of a Solr Invenio adapter should have highest priority. Xapian offers interesting capabilities and might increase performance in comparison to Solr. Its lack of multi value documents might require extra work. As a consequence, the implementation of a Xapian Invenio adapter should have second priority. elasticsearch offers very nice and flexible data format capabilities. Due to the lack of sufficient documentation, the implementation of an elasticsearch Invenio adapter should have least priority.

## 3.4 Existing integration

Besides metadata fields like title, author, abstract and collection, Invenio offers fulltext which can be extracted from record attachments. Currently, Invenio offers Solr-based indexing and search for **fulltext data**. The existing solution uses solrpy for indexing. Since this initial integration works very well, Solr seems to be promising for the integration of further fields and for ranking use cases.

Internally, Invenio uses the data structure `intbitset`. It holds an unordered set of unsigned integers. `intbitset` offers highly optimized set operations which are implemented as bit vectors and uses Python C extensions to increase performance. Using compression, it also reduces memory consumption. The existing Solr integration returns `intbitset` responses which contain the ids of the records that match the query. The Solr `intbitset` support is implemented in numerous Java classes.

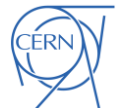

Figure 18: Solr-based fulltext search

# 4 Design

This chapter presents to which extend the concerned Invenio parts are affected. The focus is on a generic bridge with a Solr and Xapian adapter.

## 4.1 Selected information retrieval systems

Solr and Xapian adapters shall be implemented for the ranking bridge. Both information retrieval systems have proven in chapter 3 to be useable.

## 4.2 Supported fields

Both adapters shall support ranking for the following record fields initially, since they are most likely for word similarity ranking use cases:

- `Abstract`
- `Author`, concatenation of:
    - `First Author`
    - `Additional Authors`
- `Fulltext`
- `Keyword`
- `Title`

## 4.3 Searching and ranking workflow

The Invenio architecture is highly modular concerning searching and ranking. Searching and ranking are performed independently:

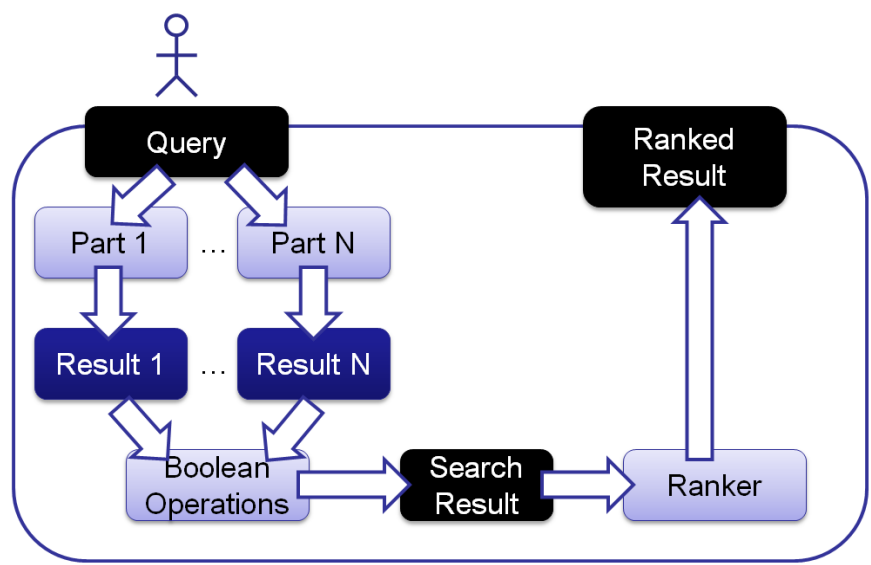

**Figure 19: Existing high-level searching and ranking workflow**

A query can contain search in different fields like `"abstract:Higgs AND title:Spin"` to search "Higgs" in the abstract field and "Spin" in the title field. Invenio searches independently per field (note that field and index are synonyms from a pure Invenio perspective). Each search returns an `intbitset` vector containing the record ids of the matches. Subsequently a boolean operation depending on the query is performed on the vectors. At this stage the result is not ranked. In the next step, the result set is ranked by passing the query and the result ids to the ranker.

Therefore, the project will mostly affect the second stage by using the generic bridge to retrieve the ranked results:

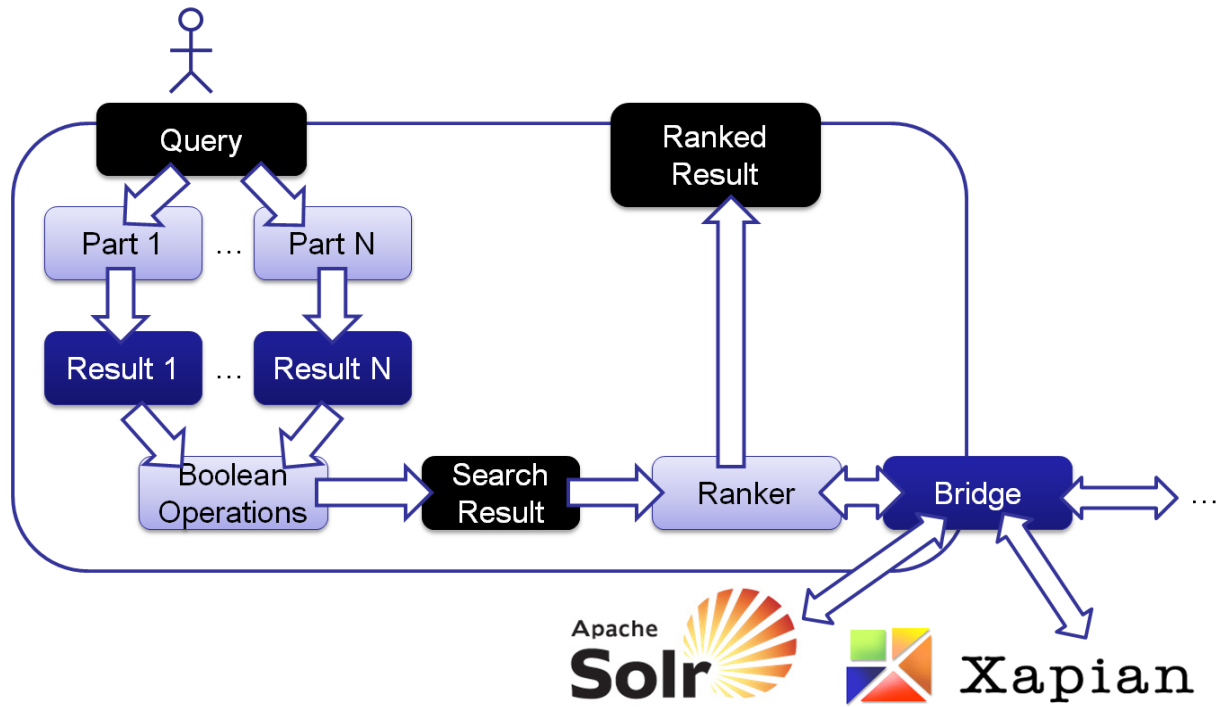

**Figure 20: Future high-level searching and ranking workflow**

## 4.4 Existing Solr integration

Currently, the existing `intbitset` Java classes support search results but no ranking results. Their key purpose is to increase performance. Since performance tuning is a non-functional requirement, no project resources will be allocated to improved Solr Java classes supporting `intbitset` for ranking.

### 4.4.1 solrutils.py

`solrutils.py` is a file containing methods which offer most of the existing Solr fulltext support. All future Solr functionality must be added to this file whenever possible.

### 4.4.2 Required amendments

`BibIndex` indexes data only for search purposes. Therefore, the current fulltext search data is indexed by `BibIndex`. Ideally, the corresponding `BibIndex` parts should direct to methods within `solrutils.py`. Unfortunately, the source code using solrpy for fulltext search is entirely within the `BibIndex` source code. It must be refactored and moved to `solrutils.py`.

## 4.5 BibRank

`BibRank` is the most-effected part of Invenio since the project focuses on ranking.

### 4.5.1 Ranking method selection

The different ranking methods can be configured within the `BibRank` admin interface including the word similarity method:

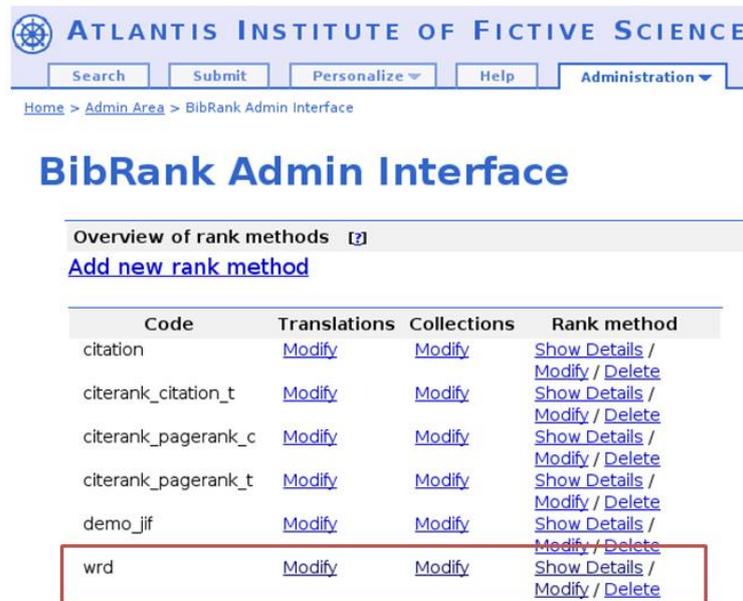

**Figure 21: BibRank admin interface**

Within the admin interface the word similarity ranking method can be configured. The interface must offer a *template* per adapter containing the function name, i.e. `template_word_similarity_solr.cfg` respective `template_word_similarity_xapian.cfg` and optional configuration parameters.

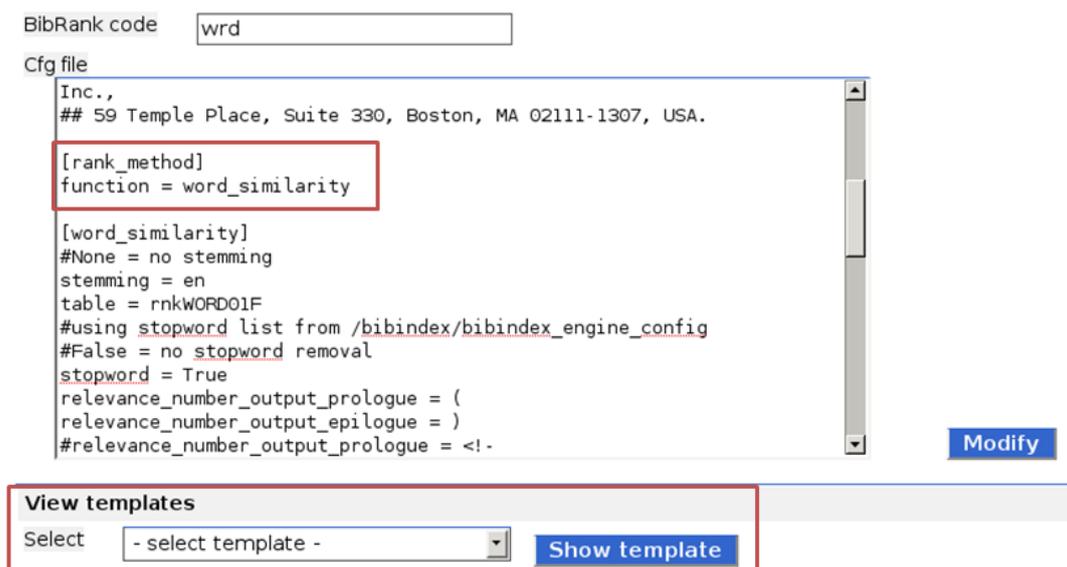

**Figure 22: Configuration of the existing word similarity ranking method**

The `BibRank` engine and all ranking methods consist of two parts: **indexing** of relevant data for ranking purposes and **searching** for ranking purposes. The **indexing part** is executed as a task within `BibSched` to pre-calculate as many data as possible as explained in chapter 2.3. To rank a result set, the **searching part** is executed using the pre-calculated data for performance reasons.

### 4.5.2 Existing word similarity searcher

All ranking methods like *citation count* except of word similarity are split into two files: `bibrank_<method>_indexer.py` and `bibrank_<method>_searcher.py`.

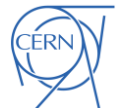

The existing native word similarity ranking method can be split better. The indexing part is implemented in `bibrank_word_indexer.py` whereas the searching part is contained within the method-independent ranking engine in `bibrank_record_sorter.py`. Since word similarity ranking is the first ranking method offered by Invenio, this might explain this monolithic structure.

The word similarity ranking searching part should be separated from the ranking engine and moved to a new file called `bibrank_word_`**`searcher`**`.py`.

### 4.5.3 Indexing

`BibRank` is run as a task within `BibSched` to index all relevant data for ranking. Currently, `BibRank` uses an in-house solution for the word similarity method. The `BibRank` indexing part must be extended to direct to `sulrutils.py` respective `xapianutils.py` when the selected information retrieval system has been set as word similarity ranking method.

Within `<adapter>utils.py`, the fields described in chapter 4.2 of all records must be added to the selected information retrieval system. It will then index the corresponding data internally making it ready to be queried efficiently.

### 4.5.4 Searching

`BibRank` receives a search result stored in an `intbitset` and the user query. The `intbitset` was calculated by `WebSearch` using `BibIndex` and contains the matched record ids. Currently, `BibRank` executes the query and subsequently receives a ranked result set from an in-house word similarity method implementation.

The `BibRank` searching part must be extended to direct to `<adapter>utils.py` when the selected information retrieval system is set as word similarity ranking method. `<adapter>utils.py` must send the query to it and subsequently receive record ids and corresponding scores. Comparing the ranking response to the original result ids might lead to two different cases:

1. Ideally, both results contain exactly the same records ids. In this case, the ranked result must be returned to `WebSearch` directly which displays the ranked result.
2. Since the selected information retrieval system and the search engine might be configured differently due to stemming, stopwords etc. both result sets might be different:
   - The ranked result contains additional records ids: these ids shall be dropped since they are not in the original result set.
   - The ranked result contains less record ids: the record ids not contained in the ranked result must be part of the eventual result but not ranked in this case.

Eventually, the modified ranked result set must be returned to `WebSearch` which displays it.

Figure 23: Ranked search result containing not ranked records

Furthermore, the basic Invenio query syntax shall be supported by the adapters like the following examples:

- `abstract:Higgs AND title:Spin`
- `abstract:Higgs OR title:Spin AND fulltext:Test`
- `abstract:Higgs OR title:Spin NOT fulltext:"Test at CERN"`

The Boolean expressions can be expressed equivalently in Invenio:

- `abstract:Higgs + title:Spin`
- `abstract:Higgs | title:Spin + fulltext:Test`
- `abstract:Higgs | title:Spin - fulltext:"Test at CERN"`

A query might contain fields that are not supported like "`abstract:Higgs AND year:2002`" where `year` is not supported. In that case, a **default field** shall be used as substitute. The ideal place for the default field definition is the template configuration file. The fulltext field is an adequate default substitute since all relevant data is saved in there if the record contains text document attachments.

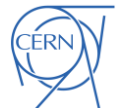

# 5 Implementation

This chapter presents the implementation aspects of the required features and amendments discussed in chapter 4. Furthermore, it presents the information retrieval system adapter implementations for Solr and Xapian.

## 5.1 Amendment of existing functionality

As discussed in chapter 4, the existing Solr fulltext search needs to be amended. Also, the existing in-house word similarity ranking implementation is split better even though it is not related directly to the Solr integration.

### 5.1.1 Solr

The existing Solr integration contains a field type definition called `invenioText` in `schema.xml` defining Invenio-specific field settings like stemming. It also contains a field called `fulltext` of this type.

The existing Python Solr fulltext indexing code using solrpy is entirely in `BibIndex`. Subsequently, it is now moved to `solrutils.py`.

### 5.1.2 Word similarity ranking

As described in chapter 4, the ranking searching part of the existing in-house word similarity ranking method is merged with the `BibRank` engine. The concerned eight functions are now moved to the newly created file `bibrank_word_searcher.py` making it consistent to the other ranking methods. This move resulted in a minimal amendment of imports and unit tests.

## 5.2 Common structure of adapters

Each adapter consists of three parts:

- A configuration template file
- A ranking indexing API
- A ranking search API

In this chapter, they are presented in more detail from an information retrieval system independent perspective.

### 5.2.1 Configuration template file

The configuration template file called `template_word_similarity_<adapter>.cfg` is related to `template_word.cfg`. It contains settings to use the in-house word similarity ranking implementation. The template file structure is the following:

```
[rank_method]
function = word_similarity_<adapter>

[...]

[field_settings]
fields = {  "abstract": {"weight":INTEGER},
            "author": {"weight":INTEGER },
            "fulltext": {"weight":INTEGER },
            "keyword": {"weight":INTEGER },
            "title": {"weight":INTEGER }
            }
default_field = fulltext
```

*Source code 16: Configuration template file structure*

First, it contains the section `[rank_method]` having an unique function name called `word_similarity_<adapter>` which is used by the ranking engine (see chapter 5.3).

Second, it contains the section `[field_settings]` which contains a *Python dictionary* called `fields` defining the supported fields and their respective properties. At this point, only *ranking weights* are defined, but further properties can be added easily. Ranking weights allow increasing the importance of fields depending on a specific use case. Since Invenio supports fulltext search, it is likely that `fulltext` has a greater weight than `title` in most use cases. It also contains a `default_field` setting in case the query contains non-supported fields. In this case, `default_field` is used as a substitute for ranking. At a first glance, a Python dictionary data structure for this part might irritate and might easily lead to security breaches. Since the configuration is done by administrators and not public available, this is not a problem. Another benefit of the Python dictionary is that it can be easily used by the ranking engine (see chapter 5.3) which does not require any sophisticated file parsing.

All configuration templates need to be stored in `/opt/invenio/etc/bibrank/`. This folder contains also a file named `wrd.cfg`. It contains the information retrieval selection and configuration and is used by the ranking engine. The existing admin interface allows editing `wrd.cfg`. Furthermore, a specific `template_word_similarity_<adapter>.cfg` can be easily copied into `wrd.cfg` by command line to select an information retrieval system.

### 5.2.2 Ranking indexing and search APIs

The ranking indexing API is in `<adapter>utils_bibrank_indexer.py` and the search API is in `<adapter>utils_bibrank_searcher.py`. The separation makes the adapter API more modular and enforces a *separation of concerns*.

The indexing API offers functionality to add the relevant record fields to the corresponding information retrieval system. It needs to offer at least one function named `word_similarity_<adapter>` which is called by the `BibRank` engine for the selected information retrieval system (see chapter 5.3.2). Ideally, this method should add status messages to the `BibRank` command line output.

The search API offers functionality to rank a search result. It also needs to offer at least one function named `word_similarity_<adapter>` which is called by the `BibRank` engine for the selected information retrieval system (see chapter 5.3.3). This method must receive the search result set and the original query to rank the result set using the indexed data in the selected information retrieval system.

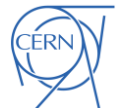

## 5.3 BibRank engine

Using the selected adapter in `wrd.cfg` requires three parts of the `BibRank` engine to be changed:

- Configuration loading
- Indexing part
- Searching part

In this chapter, the necessary amendments are presented in more detail. Furthermore, it highlights some of the *Python key strengths making the implementation shorter*.

### 5.3.1 Configuration loading

Ranking configuration files like `wrd.cfg` for `word similarity ranking` are loaded within a function named `create_rnkmethod_cache`. The new section `[field_settings]` and its corresponding fields can be loaded with very little effort. Since `fields` is represented as a Python dictionary in the configuration text file, it can be converted to a dictionary by using `eval`. The `eval` function exists in many script languages and interprets a string as a programming language expression.

```
if config.has_section("field_settings"):
    methods[…]["fields"] = eval(config.get("field_settings", "fields"))
    methods[…]["default_field"] = config.get("field_settings",
                                             "default_field")
```

<p align="center">Source code 17: Ranking configuration loading</p>

### 5.3.2 Indexing

When `BibRank` is run as a task within `BibSched`, `bibrank.py` reads the indexing function name for the word similarity configuration file `wrd.cfg`. The name is stored in the field `function` within the section `[rank_method]`. Subsequently it is called. This approach is already highly generic:

```
[…]
cfg_function = config.get("rank_method", "function")
func_object = globals().get(cfg_function)
if func_object:
    func_object(key)
[…]
```

<p align="center">Source code 18: Indexer execution within BibRank</p>

The only necessary addition to `bibrank.py` is a sequence of imports concerning all adapter indexing API functions:

```
from invenio.solrutils_bibrank_indexer import word_similarity_solr
from invenio.xapianutils_bibrank_indexer import word_similarity_xapian
```

<p align="center">Source code 19: Necessary adapter indexing API function imports</p>

The imported functions are not used directly in the source code but are necessary for the execution of a selected adapter indexer. Therefore, the static analysis tool *pylint* warns about them. The warnings can be easily suppressed by adding `#@UnusedImport` behind each import.

### 5.3.3 Searching

The searching part of the `BibRank` engine using the function name for the word similarity configuration file `wrd.cfg` is less generic. Making it highly generic would result in a complete refactoring of the searching part. Therefore, the amended `BibRank` engine simply contains another two if statements directing to the adapter search APIs:

```
[…]
elif func_object:
        if function == "word_similarity":
            result = func_object(rank_method_code, pattern, hitset, […])

        elif function == "word_similarity_solr":
            result = word_similarity_solr(pattern, hitset, […])

        elif function == "word_similarity_xapian":
            result = word_similarity_xapian(pattern, hitset, […])

        […]
```

<div align="center">Source code 20: Searcher execution within BibRank</div>

## 5.4 Solr adapter

This chapter describes the Solr adapter implementation and the adapter-specific issues that occurred.

### 5.4.1 Adapter source code files and split of solrutils.py

The Solr adapter API consists of the following files in accordance to the common adapter structure presented in chapter 5.2:

- `solrutils_bibrank_indexer.py`
- `solrutils_bibrank_searcher.py`

Since the Solr integration also contains fulltext searching and indexing, the previous `solrutils.py` is split into the following files for consistence:

- `solrutils_bibrank_indexer.py`
- `solrutils_bibrank_searcher.py`

These files can be easily extended in future with further searching and indexing functionality.

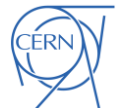

### 5.4.2 Field definitions

The supported fields are added to the pre-existing definitions (`fulltext`, `invenioText`) in the `schema.xml` file of the Solr instance.

```
[…]
<field name="fulltext" type="invenioText"  […] />
[…]
<field name="abstract" type="invenioText"  […] />
<field name="author" type="invenioText"  […] />
<field name="keyword" type="invenioText"  […] />
<field name="title" type="invenioText"  […] />
[…]
```

Source code 21: schema.xml field definitions

### 5.4.3 Indexing

The indexing API is called by `BibRank` when it is run as a task within `BibSched`. Subsequently, it adds the fields `author`, `abstract`, `keyword` and `title` of all records to Solr. See appendix E.1 for the indexing algorithm in more detail.

Usually, `BibIndex` adding all fulltext data is usually run before. Therefore, the ranking indexing API *should* update the existing documents containing the fulltext. Unfortunately, Solr does not offer updates yet. As a consequence, Solr needs to override the old documents with the new ones containing the additional fields and using the saved extracted fulltext (saved by `BibIndex`):

```
def solr_add_all(max_recid):
    for recid in range(1, max_recid + 1):
        author = […]
        abstract = […]
        keyword = […]
        title = […]
        fulltext = […] # Reads saved extracted fulltext
        solr_add(recid, abstract, author, fulltext, keyword, title)
```

Source code 22: Solr indexing preserving fulltext

### 5.4.4 Searching

The search API is called by `WebSearch` passing the original search query and an `intbitset` containing the search record result ids to be ranked called *hitset*. Subsequently, the adapter uses Solr to rank the result set.

The search algorithm consists of the following steps:

1. The Invenio search query is transformed to a Solr query
2. The query is sent to Solr
3. The ranked result scores are normalized
4. Unranked search results records are added to the ranked result
5. The ranked result elements are sorted by the score
6. The ranked result is returned to `WebSearch`

See appendix E.2 for the search algorithm in more detail.

First, the Invenio search query is transformed to a Solr query. The generated Solr query also contains weights. The following example illustrates the transformation with **fulltext** and **title** having the **weights 10** and **2** and **fulltext** being the **default ranking field**. Weights are added to `"query_part^weight"` in Solr query syntax.

**Search:**

| `"FOR NUCLEAR" + title:at - year:2010` | | `fulltext` | ▼ |

<p align="center">Figure 24: Sample Invenio query to be transformed</p>

The Solr query is composed iteratively:

- `"FOR NUCLEAR"` does not have an explicit field definition, therefore the implicit field definition `fulltext` is used and converted to: **`fulltext:"FOR NUCLEAR"^10`**
- `"+"` is converted to **`AND`**
- `"at"` has an explicit field definition, it is converted to: **`title:at^2`**
- `"-"` is converted to **`NOT`**
- `"2010"` does have the explicit non-supported field definition `year`, therefore the default ranking field `fulltext` is used and converted to: **`fulltext:2010^10`**

Subsequently, the generated Solr query is: **`fulltext:"FOR NUCLEAR"^10 AND title:at^2 NOT fulltext:2010^10`**.

Second, the query is sent to Solr. Solr returns a ranked record id result set. The result might contain record ids that are not contained in the original search result. Therefore, additional Solr results are dropped:

```python
def solr_get_ranked(query, hitset):
    """
    Sends a query to Solr.

    Returns: a list of ranked record ids [(recid, score), ...) contained in
     hitset and an intbitset of record ids contained in hitset.
    """
    response = SOLR_CONNECTION.query(q=query, fields=['id', 'score'])
    result = []
    matched_recs = intbitset()

    for hit in response.results:
        recid = int(hit['id'])
        if recid in hitset:
            score = int(float(hit['score']) * 100) + 1
            result.append((recid, score))
            matched_recs.add(recid)
[…]
```

<p align="center">Source code 23: Filtering the Solr result</p>

The scores might not look adequate since they are calculated with the vector space model. Therefore, they are normalized to the interval [0, 100] in accordance to the in-house word similarity ranking solution. Some record ids of the search result might not be returned by Solr due to different

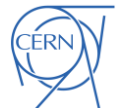

configuration like stemming. Therefore they are added to the ranking result having a 0 score. Next, they are sorted by their score. Sixth, the ranked result is returned to `WebSearch`:

```python
(ranked_result, matched_recs) = solr_get_ranked(query, hitset)
ranked_result = get_normalised_ranking_scores(ranked_result)

# Considers not ranked records
not_ranked = hitset.difference(matched_recs)
if not_ranked:
        lrecIDs = list(not_ranked)
        ranked_result = zip(lrecIDs, [0] * len(lrecIDs)) + ranked_result

ranked_result.sort(lambda x, y: cmp(x[1], y[1]))
return (ranked_result, […])
```

**Source code 24: Processing of the Solr result**

The algorithms use `intbitset` as far as possible to improve the performance of set operations like containment and difference.

## 5.5 Xapian adapter

This chapter describes the Xapian adapter implementation and the adapter-specific issues that occurred.

### 5.5.1 Adapter source code files

The Xapian adapter API consist of the following files in accordance to the common adapter structure presented in chapter 5.2:

- `xapianutils_bibrank_indexer.py`
- `xapianutils_bibrank_searcher.py`

In addition, it also includes a file called `xapianutils_config.py`. It contains configuration variables that are used by both the indexer and searcher:

```python
from invenio.config import CFG_CACHEDIR

INDEXES = ("abstract", "author", "fulltext", "keyword", "title")
DATABASES = dict()
XAPIAN_DIR_NAME = "xapian_indexes"
XAPIAN_DIR = CFG_CACHEDIR + "/" + XAPIAN_DIR_NAME
```

**Source code 25: xapianutils_config.py**

### 5.5.2 Field definitions

As described in chapter 3.3.3, Xapian does not support multiple fields per document since it is a pure string document store. Therefore, one Xapian database is created per field. This results in manual query processing as described in chapter 5.5.4.

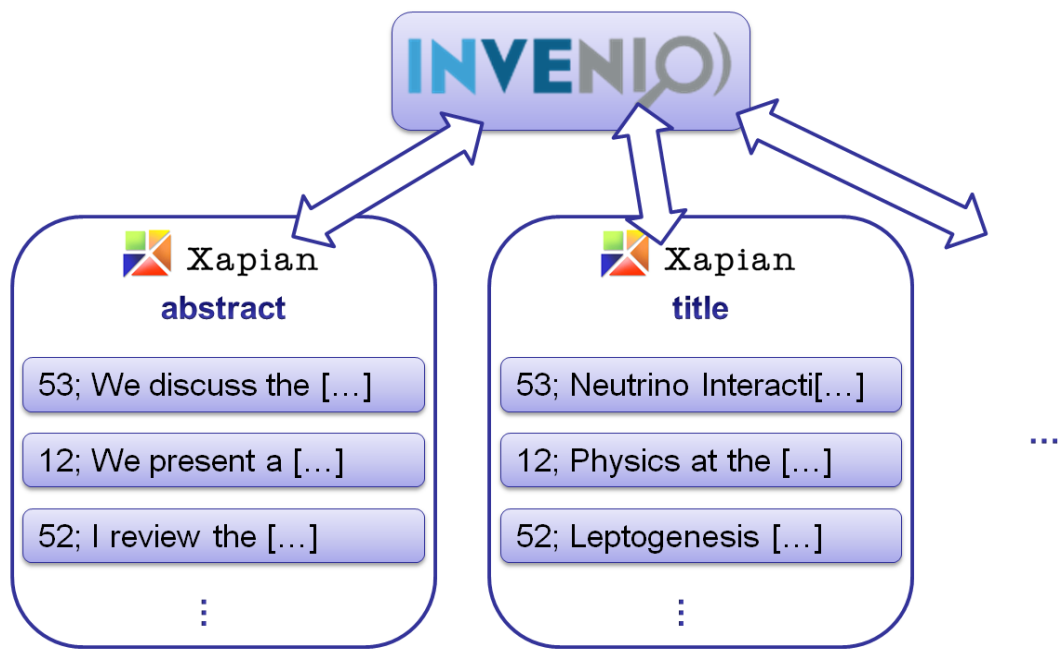

If not existent, the databases are created in the cache folder of the Invenio installation. It creates a subfolder called "xapian_indexes" and creates if not existent one subfolder per field within it containing the respective database. It also stores a reference to the database object and the corresponding indexer object (see chapter 3.3.3) in the global dictionary `DATABASES`:

```
def xapian_init_databases():
    xapian_ensure_db_dir(XAPIAN_DIR_NAME)
    for field in INDEXES:
        xapian_ensure_db_dir(XAPIAN_DIR_NAME + "/" + field)
        database = xapian.WritableDatabase(XAPIAN_DIR + "/" + field,
                                           xapian.DB_CREATE_OR_OPEN)
        indexer = xapian.TermGenerator()
        stemmer = xapian.Stem("english")
        indexer.set_stemmer(stemmer)
        DATABASES[field] = (database, indexer)
```

Source code 26: Xapian databases creation

All settings like stemming are done in the searching and indexing source code as presented in chapter 3.3.3.

### 5.5.3 Indexing

Similar to the Solr adapter, the indexing API adds the fields `author`, `abstract`, `keyword` and `title` of all records to Xapian when it is called by a `BibRank` task. See appendix E.3 for the indexing algorithm in more detail.

First, the databases, the database objects and indexer objects need to be created as presented in chapter 5.5.2. Second, the saved fulltext is added to the Xapian databases in conjunction with the other fields per record.

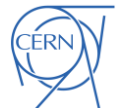

```python
def xapian_add_all(max_recid):
    xapian_init_databases()
    for recid in range(1, max_recid + 1):
        abstract = […]
        xapian_add(recid, "abstract", abstract)

        author = […]
        xapian_add(recid, "author", author)

        fulltext = […] # Reads saved extracted fulltext
        xapian_add(recid, "fulltext", fulltext)

        keyword = […]
        xapian_add(recid, "keyword", keyword)

        title = […]
        xapian_add(recid, "title", title)
```

<div align="center">Source code 27: Xapian indexing</div>

### 5.5.4 Searching

Similar to the Solr adapter, the search API is called by `WebSearch` and receives the query and an `intbitset` called *hitset*. It contains the search record result ids to be ranked.

The search algorithm consists of the following steps:

1. The Invenio query is executed per query part and the results are aggregated
2. The excluded records are removed (`"-" something`)
3. The result is filtered by the greatest score per record
4. The ranked scores are normalized
5. Unranked search results are added to the ranked result
6. The ranked result elements are sorted by the score
7. The ranked result is returned to `WebSearch`

See appendix E.4 for the search algorithm in more detail.

Since one Xapian database is created per field, complex queries consisting of more than one part need to be split up. The results are combined in the source code. This might result in a complex query processing considering operator precedence. Therefore, operator evaluation is simplified for ranking which does not affect the search result itself:

- Both "+" and "|" are interpreted as "OR"
- "-" is interpreted as "NOT"

First, all query parts are executed and "+" and "|" parts are aggregated. Excluded parts ("-") are aggregated in a separate data structure. A default ranking field is used the same way as in the Solr adapter. Weights are multiplied by relevance score returned by Xapian per query part. This shall be demonstrated by the following example (all record ids and ranking scores are artificial):

```
"FOR NUCLEAR" | title:neutrino - year:2002 + title:at        fulltext ▼
```

<div align="center">Figure 26: Sample Invenio query to be evaluated</div>

It is evaluated part-wise (syntax is `(record_id, score)`):

1. "`FOR NUCLEAR`" does not have an explicit field definition, therefore the implicit field definition `fulltext` is used. Xapian returns the following result: `[(3, 10), (31, 20)]` which is added to `all_ranked_results`.

2. Xapian returns for the query "`| title:neutrino`" the following result: `[(31, 40), (20, 30), (4, 5)]` which is added to `all_ranked_results`. `all_ranked_results` is now `[(3, 10), (31, 20), (31, 40), (20, 30), (4, 5)]`.

3. "`2002`" does have the non-support explicit field definition `year`, therefore the default ranking field `fulltext` is used. Xapian returns for the query "`- fulltext:2002`" the following result: `[(3, 30), (7, 2)]`. Since these results shall be excluded, only their record ids are added to `excluded_hits`.

4. Xapian returns for the query "`+ title:at`" the following result: `[(7, 5)]` which is added to `all_ranked_results`. `all_ranked_results` is now `[(3, 10), (31, 20), (31, 40), (20, 30), (4, 5), (7, 5)]`.

Second, the excluded hits `[3, 7]` are removed from `all_ranked_result` which is now `[(31, 20), (31, 40), (20, 30), (4, 5)]`. At this stage, records might be in `all_ranked_result` for multiple times. Third, it is filtered by the greatest score per record. Therefore, `all_ranked_result` is now `[(31, 40), (20, 30), (4, 5)]`:

```python
def get_greatest_ranked_records(raw_reclist):
    unique_records = dict()
    for (recid, score) in raw_reclist:
        if not unique_records.has_key(recid):
            unique_records[recid] = score
        else:
            current_score = unique_records[recid]
            if score > current_score:
                unique_records[recid] = score

    result = []
    for recid in unique_records.keys():
        result.append((recid, unique_records[recid]))

    return result
```

Source code 28: Filter result records by greatest score

Steps four to seven are equivalent to the Solr adapter in chapter 5.4.4: The scores might not look adequate since they are calculated with the vector space model. Therefore, they are normalized to the interval [0, 100] in accordance to the in-house word similarity ranking solution. Some record ids of the search result might not be returned by Xapian due to different configuration like stemming etc. Therefore they are added to the ranking result having a 0 score. Next, they are sorted by their score. Last, the ranked result is returned to `WebSearch`.

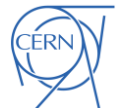

# 6 Scalability

In this chapter, the Solr adapter is evaluated in a scalability study. Different factors including indexing time, index size and query performance are measured for a function of growing amount of data. Multi-user query execution scalability is not evaluated.

## 6.1 System and configuration

The test system is a Dell PowerEdge M600 0MY736 server. It has two Intel Xeon E5410 CPUs @ 2.33GHz and eight cores in total. Furthermore, it contains 16 GB RAM and two SCSI hard disks with 146 GB each. It uses Scientific Linux CERN 5 (SLC5) [SL5] as operating system.

The system uses Invenio, MySQL, Apache etc. configuration that is used and recommended for production servers [SC5]. Apache is disabled for the following tests since no Invenio web interface is necessary. Solr uses exactly the same configuration as before which was created in the preliminary fulltext search project. Background jobs were reduced to a minimum but the study might be still distorted by background activity.

## 6.2 Data

This study uses CERN Document Server (CDS) [CDS] records which are mostly **high energy physics related**. A record's fulltext is the aggregated fulltext of all record attachments. The fulltext extraction was run in advance and the extracted fulltexts are stored on a different server and mounted. In total 400,000 records that are guaranteed to have attachments are used. Record attachment fulltexts might be significantly different:

- Short: short papers, news
- Midsize: long papers, BSc theses
- Long: books, PhD theses

## 6.3 Procedure

The 400,000-record set is split up in 8 parts containing 50,000 records each. The following steps are executed subsequently per part. To avoid clusters distorting the scalability study, the 400,000-record set is *shuffled*. This might not fully eliminate clusters but it is an efficient first-order approximation to reduce the likelihood of their appearance tremendously.

First, a `BibRank` task is executing adding all 50,000 records to Solr. The previously created fields `abstract`, `author`, `fulltext`, `title`, `keyword` and `title` are filled as far as possible. Second, The Solr `optimize` command is executed. It reorders an existing index structure to improve query performance and to defragment it if records were deleted [SOO].

Third, a set of Invenio **fulltext queries** is executed to return a ranked result set. Therefore, Invenio first needs to perform the request to return a *hitset* which is passed to `BibRank`. The following queries are executed:

- Term queries:
    1. `"of"`: worst case, is supposed to return a very large amount of results
    2. `"model"`: hard case, is supposed to return a large amount of results
    3. `"boson"`: usual case: is supposed to return a typical amount of results
- Phrase queries:
    1. `"of the"`: worst case, is supposed to return a very large amount of results
    2. `"phys rev"`: hard case, is supposed to return a large amount of results
    3. `"standard model"`: hard case, is supposed to return a large amount of results
    4. `"higgs boson"`: usual case, is supposed to return a typical amount of results

The Solr adapter only receives the ten most relevant results from Solr which is configured by default in `solrconfig.xml`. It then adds the not ranked ones to the result set with a 0% ranking as described in chapters 4.5 and 5.4.

In each step, the following factors are measured:

- Indexing time
- Index size
- For each query:
    - Average search query responses time (with IPython `%timeit` [IPY])
    - Result count
    - Average ranking query responses time (with IPython `%timeit`)

## 6.4 Required changes

A profiling pre-study for the first 50,000 records has shown that the current adapter spends almost 50% of its indexing time on the Solr connection. Almost another 50% are spent on loading and aggregating fulltexts.

As shown in appendix E.1, the Solr `commit` command is executed after each record insertion (function `solr_add`). It has been moved to end of the task execution and is therefore only executed once. This reduced the time spent in `solr_add` to about a fourth of its original time consumption.

The extracted fulltexts are stored on a different server and mounted. They are accessed through *CERN AFS* [AFS].

## 6.5 Results

This chapter presents the results of the scalability study and visualizes them in graphs.

### 6.5.1 Indexing time and index size

Both the indexing time and index size grow linearly for the growing amount of data:

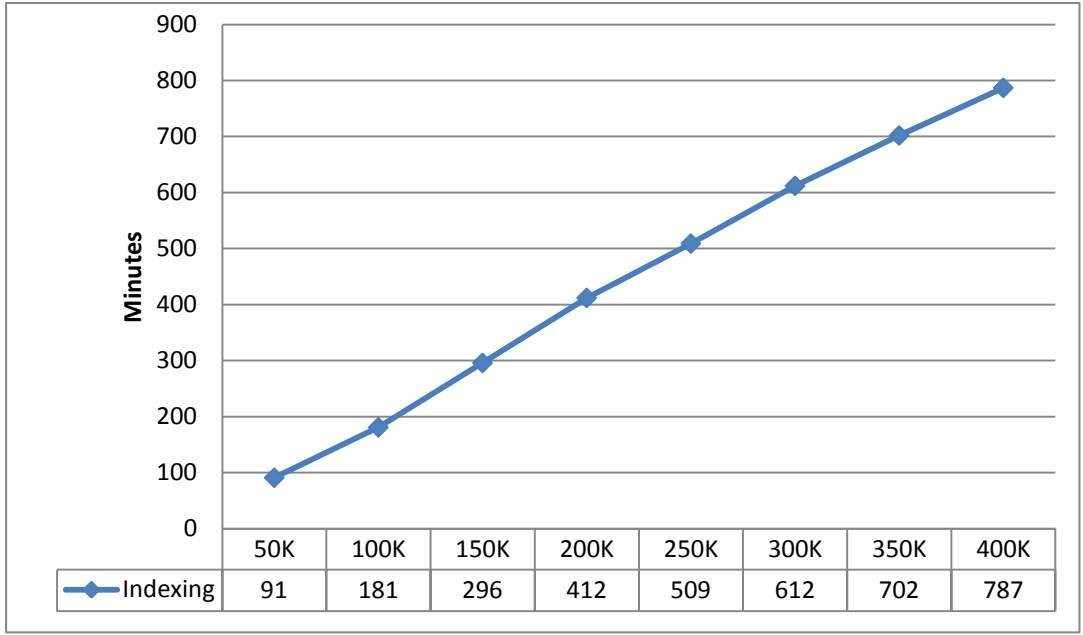

| | 50K | 100K | 150K | 200K | 250K | 300K | 350K | 400K |
|---|---|---|---|---|---|---|---|---|
| Indexing | 91 | 181 | 296 | 412 | 509 | 612 | 702 | 787 |

**Figure 27: Solr indexing time**

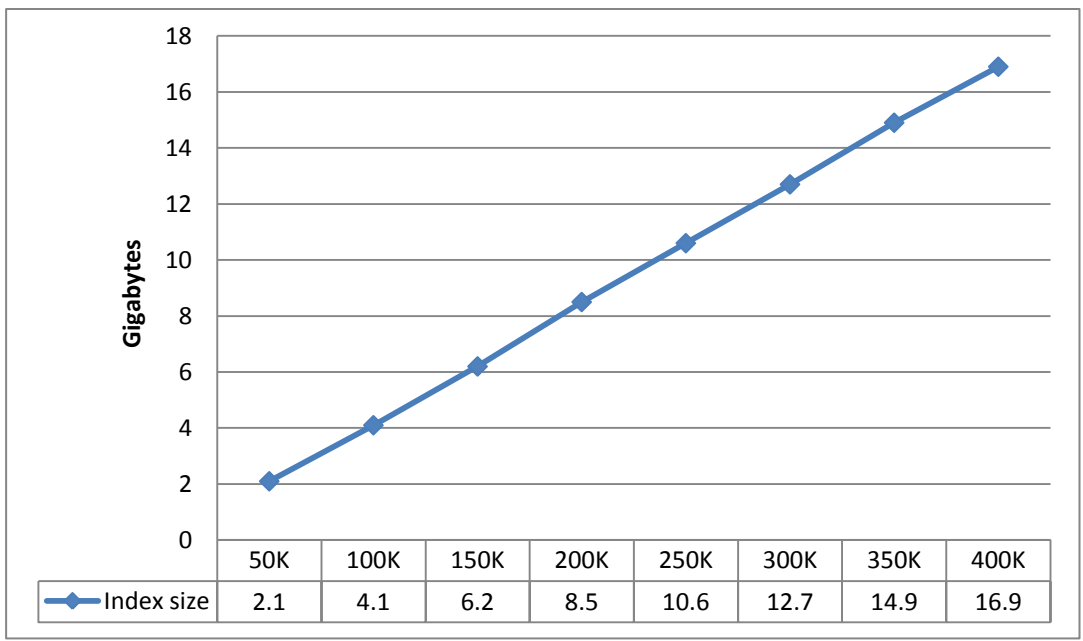

| | 50K | 100K | 150K | 200K | 250K | 300K | 350K | 400K |
|---|---|---|---|---|---|---|---|---|
| Index size | 2.1 | 4.1 | 6.2 | 8.5 | 10.6 | 12.7 | 14.9 | 16.9 |

**Figure 28: Solr index size**

### 6.5.2 Search

The queries return different amounts of results as predicted and classified in chapter 6.3:

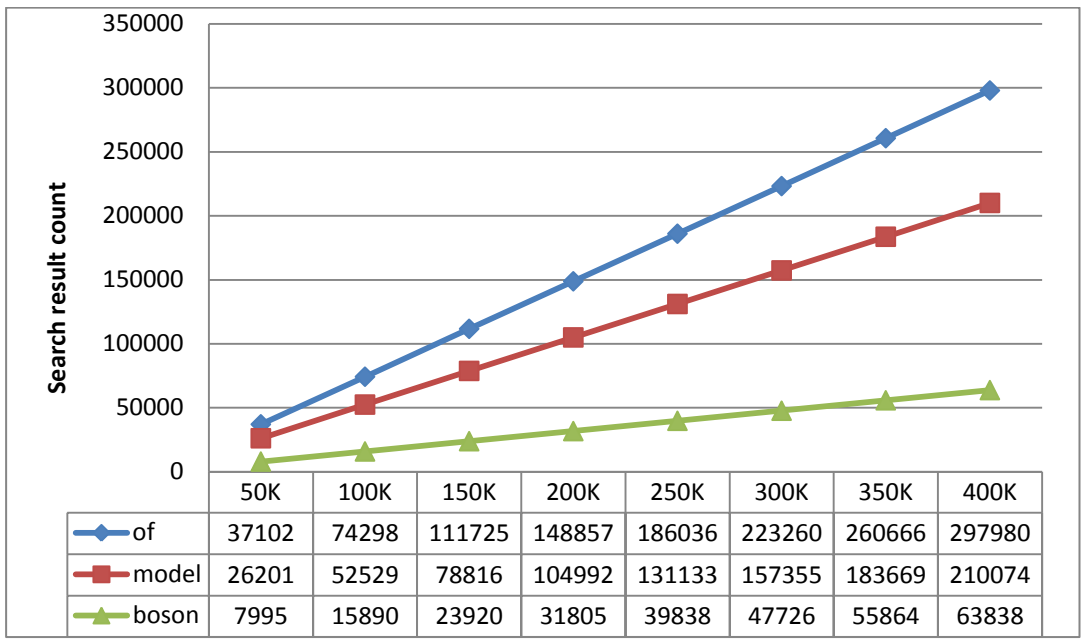

**Figure 29: Solr term queries search result count**

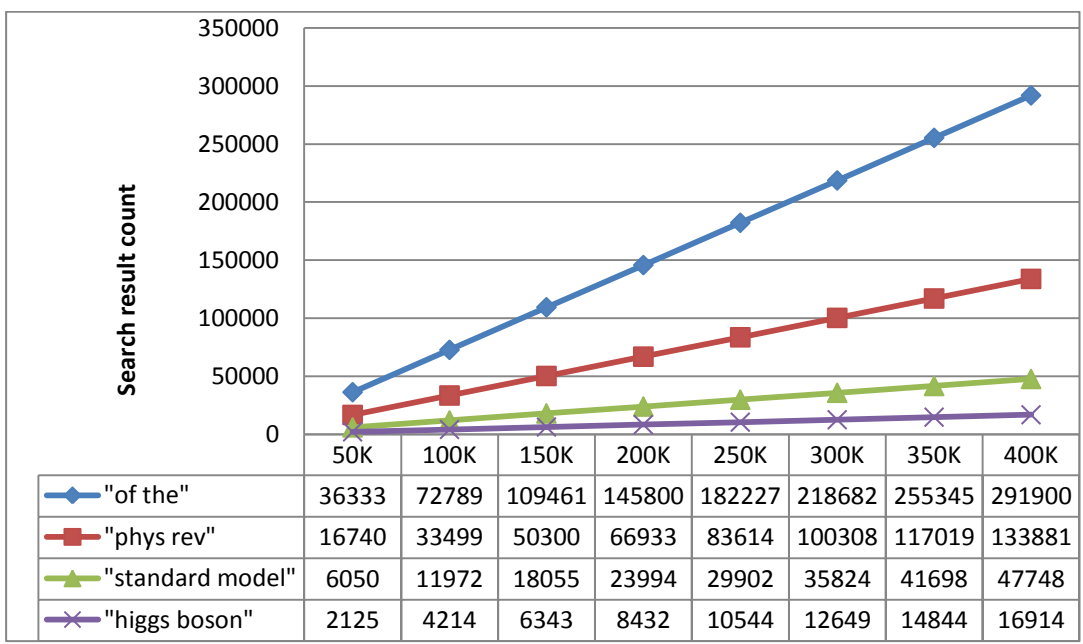

**Figure 30: Solr phrase queries search result count**

The term search queries scale well and are approximately linear:

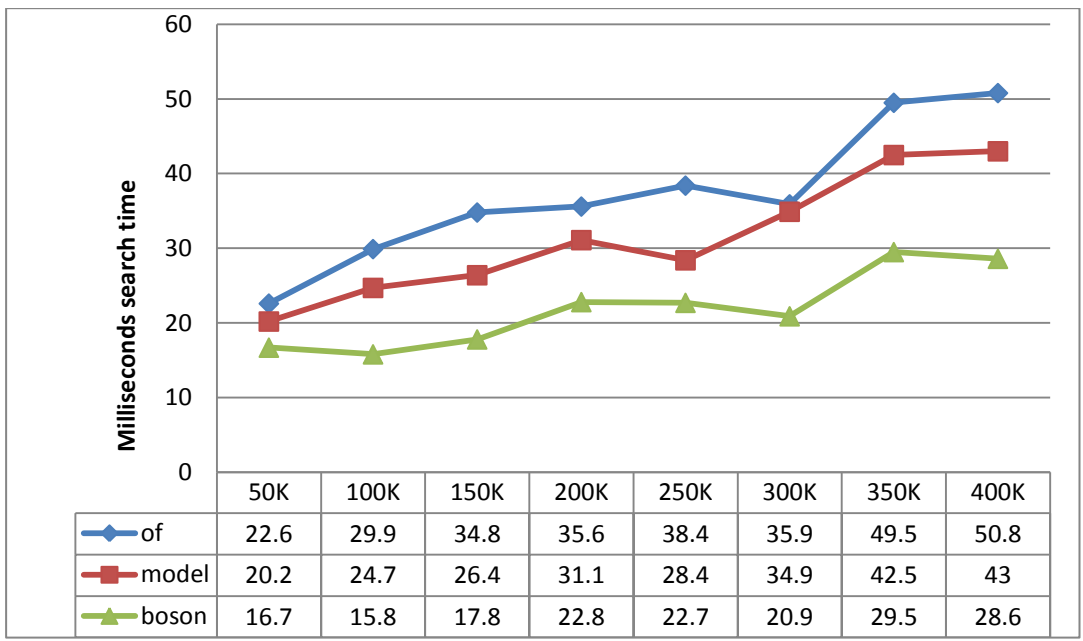

| | 50K | 100K | 150K | 200K | 250K | 300K | 350K | 400K |
|---|---|---|---|---|---|---|---|---|
| of | 22.6 | 29.9 | 34.8 | 35.6 | 38.4 | 35.9 | 49.5 | 50.8 |
| model | 20.2 | 24.7 | 26.4 | 31.1 | 28.4 | 34.9 | 42.5 | 43 |
| boson | 16.7 | 15.8 | 17.8 | 22.8 | 22.7 | 20.9 | 29.5 | 28.6 |

Figure 31: Solr term queries search time

The phrase search queries grow linear with the rarely used "of the" query consuming an extraordinary amount of time since it is a worst case scenario:

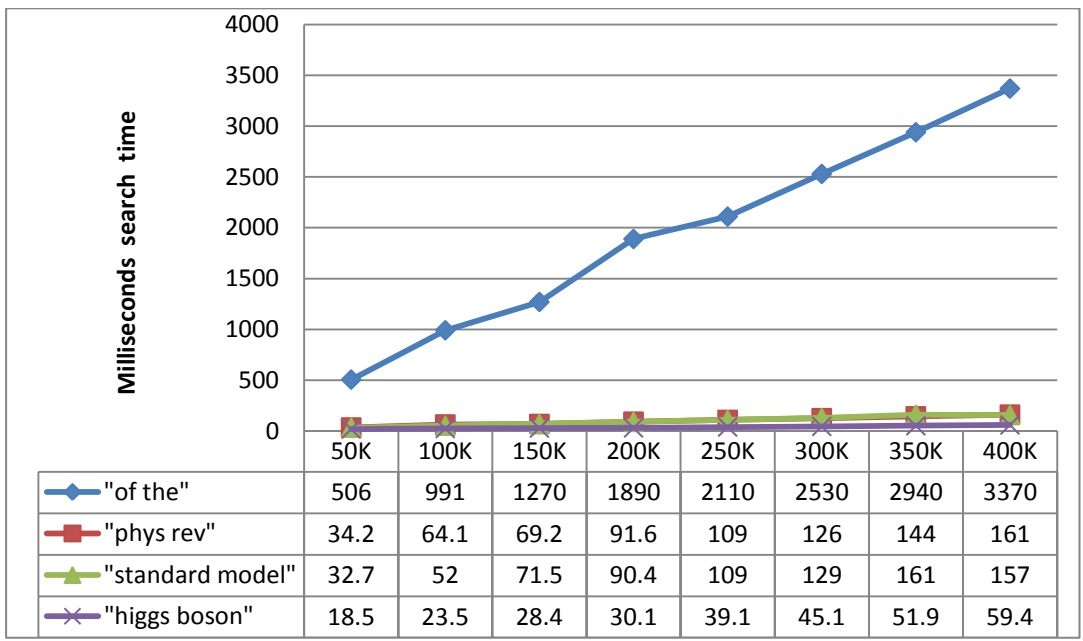

| | 50K | 100K | 150K | 200K | 250K | 300K | 350K | 400K |
|---|---|---|---|---|---|---|---|---|
| "of the" | 506 | 991 | 1270 | 1890 | 2110 | 2530 | 2940 | 3370 |
| "phys rev" | 34.2 | 64.1 | 69.2 | 91.6 | 109 | 126 | 144 | 161 |
| "standard model" | 32.7 | 52 | 71.5 | 90.4 | 109 | 129 | 161 | 157 |
| "higgs boson" | 18.5 | 23.5 | 28.4 | 30.1 | 39.1 | 45.1 | 51.9 | 59.4 |

Figure 32: Solr phrase queries search time

### 6.5.3 Ranking time

The growing amount of results is ranked in about linear time:

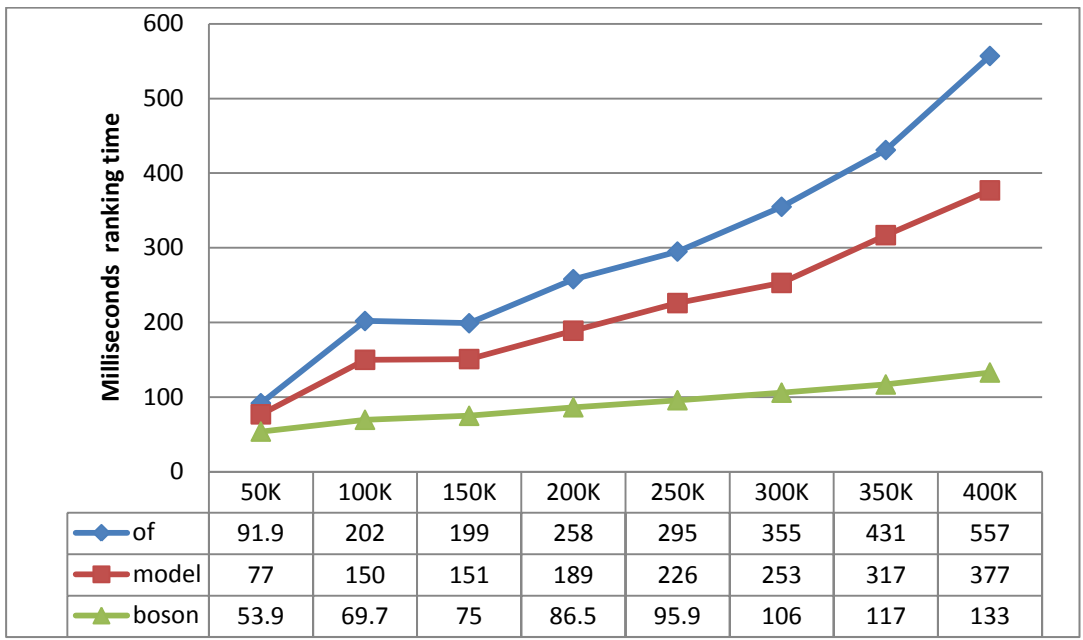

| | 50K | 100K | 150K | 200K | 250K | 300K | 350K | 400K |
|---|---|---|---|---|---|---|---|---|
| of | 91.9 | 202 | 199 | 258 | 295 | 355 | 431 | 557 |
| model | 77 | 150 | 151 | 189 | 226 | 253 | 317 | 377 |
| boson | 53.9 | 69.7 | 75 | 86.5 | 95.9 | 106 | 117 | 133 |

**Figure 33: Solr term queries ranking time**

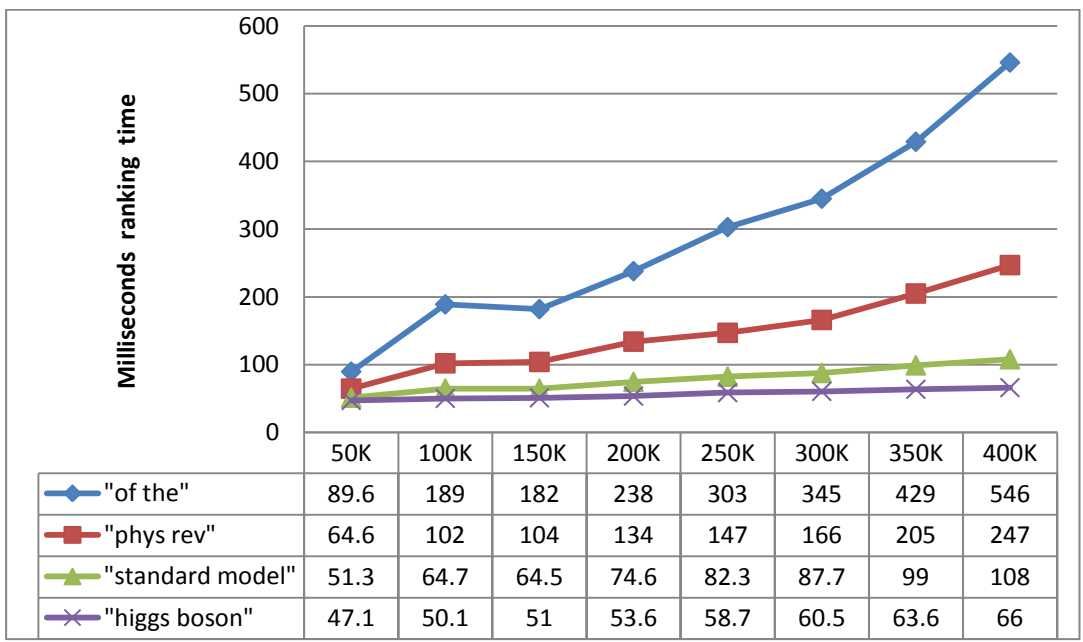

| | 50K | 100K | 150K | 200K | 250K | 300K | 350K | 400K |
|---|---|---|---|---|---|---|---|---|
| "of the" | 89.6 | 189 | 182 | 238 | 303 | 345 | 429 | 546 |
| "phys rev" | 64.6 | 102 | 104 | 134 | 147 | 166 | 205 | 247 |
| "standard model" | 51.3 | 64.7 | 64.5 | 74.6 | 82.3 | 87.7 | 99 | 108 |
| "higgs boson" | 47.1 | 50.1 | 51 | 53.6 | 58.7 | 60.5 | 63.6 | 66 |

**Figure 34: Solr phrase queries ranking time**

## 6.6 Evaluation

In this scalability study for up 400K CERN Document Server records, the performance of the Solr adapter is found good. The evaluated factors indexing time, index size and query time grow approximately linearly.

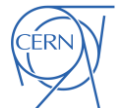

# 7 Conclusions and prospects

In this chapter, the conclusions of this project are presented followed by suggestions on how to continue the development.

## 7.1 Conclusions

The goal of this thesis was to enhance Invenio by bridging it with modern external information retrieval systems for word similarity ranking. Based on the preliminary successful fulltext search integration of Solr, the integration of external information retrieval systems was to be extended to word similarity ranking. This permits to implement enhanced ranking techniques and improves scalability.

The analysis has shown that the evaluated information retrieval systems have different strengths and weaknesses and that there is no ideal candidate to select. Both Solr and Xapian were evaluated positively.

A generic bridge for metadata indexing and word similarity ranking was designed and implemented. Both Solr and Xapian were integrated even though they work fundamentally differently. The development of both adapters has resulted in information retrieval system specific technical issues that were solved. Both adapters return reasonable results. A possibility of different weight of fields in the final ranking was introduced and implemented for both adapters. Due to the generic approach, additional information retrieval systems can be integrated in future by writing corresponding adapters.

A preliminary study of scalability was carried out for the Solr adapter for the CERN Document Server data. The indexing and searching performance was evaluated as a function of document corpus size. The performance was found good up to 400K documents with linear tendency for indexing time, index size and query time.

Both adapters have proven to be extremely powerful, they enrich Invenio very well and the respective information retrieval systems offer way more functionality that can be easily integrated in future through the generic bridge. It was very interesting to work on this project and it was especially rewarding to see it put to production tests. The developed code is going to be part of the next Invenio release and is going to be put to CERN Document Server production in the next weeks.

## 7.2 Prospects

Two parts of the implemented modules could be more generic. On the one hand side, the `BibRank` search engine requires one `if` statement per adapter. This could be avoided by refactoring the `BibRank` search engine. On the other hand side, both adapters could be refactored to use only configuration file field-wise settings for indexing and searching.

It would be useful to use both information retrieval systems not only for ranking but also for search. Especially a Solr search integration would require only a very fair amount of extra code. Nevertheless, a more precise configuration of stemming, stopwords etc. would be necessary to return results that are close to the ones returned by the current Invenio search engine.

A complementary study of performance in multi-user query conditions for both Solr and Xapian will be studied subsequent to this project.

# Appendices

## A List of figures

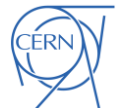

# B List of tables

# C List of source codes

# D References

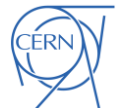

# E Source codes

## E.1 Solr ranking indexing

```python
from invenio.config import CFG_SOLR_URL
from invenio.bibtask import write_message
from invenio.dbquery import run_sql
from invenio.search_engine import get_fieldvalues
from invenio.bibsword_config import get_tag_from_name
from invenio.solrutils_bibrank_searcher import solr_get_all_fulltext

if CFG_SOLR_URL:
    import solr
    SOLR_CONNECTION = solr.SolrConnection(CFG_SOLR_URL)

def solr_add_all(max_recid):
    """
    Adds the regarding field values of all records up to max_recid to Solr.
    It preserves the fulltext information.
    """
    for recid in range(1, max_recid + 1):
        try:
            abstract = unicode(get_fieldvalues(recid,
                            get_tag_from_name("abstract"))[0], 'utf-8')
        except:
            abstract = ""
        try:
            first_author = get_fieldvalues(recid,
                            get_tag_from_name("first author name"))[0]
            additional_authors = reduce(lambda x, y: x + " " + y,
                            get_fieldvalues(recid, get_tag_from_name(
                                "additional author name")))
            author = unicode(first_author + " " + additional_authors,
                            'utf-8')
        except:
            author = ""
        try:
            bibrecdocs = BibRecDocs(recid)
            fulltext = unicode(remove_control_characters(
                                bibrecdocs.get_text()), 'utf-8')
        except:
            fulltext = ""
        try:
            keyword = unicode(get_fieldvalues(recid, get_tag_from_name(
                        "keyword"))[0], 'utf-8')
        except:
            keyword = ""
        try:
            title = unicode(get_fieldvalues(recid, get_tag_from_name(
                        "title"))[0], 'utf-8')
        except:
            title = ""
        solr_add(recid, abstract, author, fulltext, keyword, title)
```

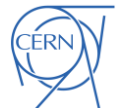

```python
def solr_add(recid, abstract, author, fulltext, keyword, title):
    """
    Helper function that adds word similarity ranking relevant indexes to
      Solr.
    """
    SOLR_CONNECTION.add(id=recid, abstract=abstract, author=author,
                        fulltext=fulltext, keyword=keyword, title=title)
    SOLR_CONNECTION.commit()

def word_similarity_solr(run):
    return word_index(run)

def word_index(run):
    """
    Runs the indexing task.
    """
    max_recid = 0
    res = run_sql("SELECT max(id) FROM bibrec")
    if res and res[0][0]:
        max_recid = int(res[0][0])
    write_message("Solr ranking indexer called")
    solr_add_all(max_recid)
    write_message("Solr ranking indexer completed")
```

### E.2 Solr ranking searching

```python
from invenio.config import CFG_SOLR_URL
from invenio.intbitset import intbitset

if CFG_SOLR_URL:
    import solr
    SOLR_CONNECTION = solr.SolrConnection(CFG_SOLR_URL)

BOOLEAN_EQUIVALENTS = {"+": "AND",
                       "|": "OR",
                       "-": "NOT"
                       }

def solr_get_ranked(query, hitset):
    """
    Sends a query to Solr.
    Returns: a list of ranked record ids [(recid, score), ...) contained in
      hitset and an intbitset of record ids contained in hitset.
    """
    response = SOLR_CONNECTION.query(q=query, fields=['id', 'score'])
    result = []
    matched_recs = intbitset()

    for hit in response.results:
        recid = int(hit['id'])
        if recid in hitset:
            score = int(float(hit['score']) * 100) + 1
            result.append((recid, score))
            matched_recs.add(recid)
    return (result, matched_recs)
```

```python
def get_normalised_ranking_scores(raw_reclist):
    """
    Returns the result having normalised ranking scores, interval [0, 100].
    """
    max_score = 0
    reclist = []

    for (recid, score) in raw_reclist:
        if score > max_score:
            max_score = score
    for (recid, score) in raw_reclist:
        normalised_score = int(100.0 / max_score * score)
        reclist.append((recid, normalised_score))

    return reclist

def word_similarity_solr(pattern, hitset, params, verbose, field):
    """
    Ranking a records containing specified words and returns a sorted list.
    [...]
    """
    voutput = ""
    search_units = []

    if pattern:
        pattern = " ".join(map(str, pattern))
        from invenio.search_engine import create_basic_search_units
        search_units = create_basic_search_units(None, pattern, field)

    query = ""
    for (operator, pattern, field, unit_type) in search_units:
        # Field might not exist
        if field not in params["fields"].keys():
            field = params["default_field"]

        if unit_type == "a":
            # Eliminates leading and trailing %
            if pattern[0] == "%":
                pattern = pattern[1:-1]
            pattern = "\"" + pattern + "\""

        weighting = "^" + str(params["fields"][field]["weight"])

        query_part = field + ":" + pattern + weighting

        # Considers boolean operator from the second part on, allows
        # negation from the first part on
        if query or operator == "-":
            query += " " + BOOLEAN_EQUIVALENTS[operator] + " "
        query += query_part + " "

    ranked_result = []

    if query:
        (ranked_result, matched_recs) = solr_get_ranked(query, hitset)
        ranked_result = get_normalised_ranking_scores(ranked_result)

        # Considers not ranked records
        not_ranked = hitset.difference(matched_recs)
        if not_ranked:
            lrecIDs = list(not_ranked)
```

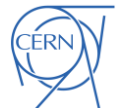

```
            ranked_result = zip(lrecIDs, [0] * len(lrecIDs))
                                        + ranked_result

        ranked_result.sort(lambda x, y: cmp(x[1], y[1]))
        return (ranked_result, params["prefix"], params["postfix"],
                voutput)

    return (ranked_result, "", "", voutput)
```

## E.3 Xapian ranking indexing

```python
import xapian
import os
from invenio.config import CFG_CACHEDIR
from invenio.bibtask import write_message
from invenio.dbquery import run_sql
from invenio.search_engine import get_fieldvalues
from invenio.bibsword_config import get_tag_from_name
from invenio.solrutils_bibrank_searcher import solr_get_all_fulltext
from invenio.xapianutils_config import DATABASES, XAPIAN_DIR, \
                                       XAPIAN_DIR_NAME, INDEXES

def xapian_ensure_db_dir(name):
    path = CFG_CACHEDIR + "/" + name
    if not os.path.exists(path):
        os.makedirs(path)

def xapian_add_all(max_recid):
    """
    Adds the regarding field values of all records up to max_recid to
      Xapian.
    It preserves the fulltext information.
    """
    xapian_init_databases()
    for recid in range(1, max_recid + 1):
        try:
            abstract = unicode(get_fieldvalues(recid,
                            get_tag_from_name("abstract"))[0], 'utf-8')
        except:
            abstract = ""
        xapian_add(recid, "abstract", abstract)

        try:
            first_author = get_fieldvalues(recid,
                            get_tag_from_name("first author name"))[0]
            additional_authors = reduce(lambda x, y: x + " " + y,
                            get_fieldvalues(recid, get_tag_from_name(
                                "additional author name")))
            author = unicode(first_author + " " + additional_authors,
                            'utf-8')
        except:
            author = ""
        xapian_add(recid, "author", author)

        try:
            bibrecdocs = BibRecDocs(recid)
            fulltext = unicode(bibrecdocs.get_text(), 'utf-8')
        except:
            fulltext = ""
        xapian_add(recid, "fulltext", fulltext)
```

```python
    try:
        keyword = unicode(get_fieldvalues(recid, get_tag_from_name(
                    "keyword"))[0], 'utf-8')
    except:
        keyword = ""
    xapian_add(recid, "keyword", keyword)

    try:
        title = unicode(get_fieldvalues(recid, get_tag_from_name(
                    "title"))[0], 'utf-8')
    except:
        title = ""
    xapian_add(recid, "title", title)

def xapian_add(recid, field, value):
    """
    Helper function that adds word similarity ranking relevant indexes to
     Solr.
    """
    content_string = value
    doc = xapian.Document()
    doc.set_data(content_string)
    (database, indexer) = DATABASES[field]
    indexer.set_document(doc)
    indexer.index_text(content_string)
    database.replace_document(recid, doc)

def xapian_init_databases():
    """
    Initializes all database objects.
    """
    xapian_ensure_db_dir(XAPIAN_DIR_NAME)
    for field in INDEXES:
        xapian_ensure_db_dir(XAPIAN_DIR_NAME + "/" + field)
        database = xapian.WritableDatabase(XAPIAN_DIR + "/" + field,
                                            xapian.DB_CREATE_OR_OPEN)
        indexer = xapian.TermGenerator()
        stemmer = xapian.Stem("english")
        indexer.set_stemmer(stemmer)
        DATABASES[field] = (database, indexer)

def word_similarity_xapian(run):
    return word_index(run)

def word_index(run):
    """
    Runs the indexing task.
    """
    max_recid = 0
    res = run_sql("SELECT max(id) FROM bibrec")
    if res and res[0][0]:
        max_recid = int(res[0][0])
    write_message("Xapian ranking indexer called")
    xapian_add_all(max_recid)
    write_message("Xapian ranking indexer completed")
```

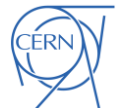

### E.4 Xapian ranking searching

```python
import xapian
from invenio.intbitset import intbitset
from invenio.dbquery import run_sql
from invenio.xapianutils_config import INDEXES, XAPIAN_DIR
from invenio.solrutils_bibrank_searcher import \
                                      get_normalised_ranking_scores

DATABASES = dict()

def xapian_get_ranked_index(index, pattern, weight, hitset, max_recid):
    """
    Queries a Xapian index.
    Returns: a list of ranked record ids [(recid, score), ...) contained in
      hitset and an intbitset of record ids contained in hitset.
    """
    result = []
    matched_recs = intbitset()

    database = DATABASES[index]
    enquire = xapian.Enquire(database)
    query_string = pattern
    qp = xapian.QueryParser()
    stemmer = xapian.Stem("english")
    qp.set_stemmer(stemmer)
    qp.set_database(database)
    qp.set_stemming_strategy(xapian.QueryParser.STEM_SOME)
    pattern = qp.parse_query(query_string, xapian.QueryParser.FLAG_PHRASE)
    enquire.set_query(pattern)
    matches = enquire.get_mset(0, max_recid)

    for match in matches:
        recid = match.docid
        if recid in hitset:
            score = int(match.percent) * weight
            result.append((recid, score))
            matched_recs.add(recid)
    return (result, matched_recs)

def xapian_init_databases():
    """
    Initializes all database objects.
    """
    for field in INDEXES:
        database = xapian.Database(XAPIAN_DIR + "/" + field)
        DATABASES[field] = database

def get_greatest_ranked_records(raw_reclist):
    """
    Returns unique records having selecting the ones with the greatest
      records in case of duplicates.
    """
    unique_records = dict()
    for (recid, score) in raw_reclist:
        if not unique_records.has_key(recid):
            unique_records[recid] = score
        else:
            current_score = unique_records[recid]
            if score > current_score:
```

```python
            unique_records[recid] = score

    result = []
    for recid in unique_records.keys():
        result.append((recid, unique_records[recid]))

    return result

def word_similarity_xapian(pattern, hitset, params, verbose, field):
    """
    Ranking a records containing specified words and returns a sorted list.
    [...]
    """
    voutput = ""
    search_units = []

    max_recid = 0
    res = run_sql("SELECT max(id) FROM bibrec")
    if res and res[0][0]:
        max_recid = int(res[0][0])

    if pattern:
        xapian_init_databases()
        pattern = " ".join(map(str, pattern))
        from invenio.search_engine import create_basic_search_units
        search_units = create_basic_search_units(None, pattern, field)

    all_ranked_results = []
    included_hits = intbitset()
    excluded_hits = intbitset()
    for (operator, pattern, field, unit_type) in search_units:
        # Field might not exist
        if field not in params["fields"].keys():
            field = params["default_field"]

        if unit_type == "a":
            # Eliminates leading and trailing %
            if pattern[0] == "%":
                pattern = pattern[1:-1]
            pattern = "\"" + pattern + "\""

        weight = params["fields"][field]["weight"]
        (ranked_result_part, matched_recs) = xapian_get_ranked_index(field,
                                pattern, weight, hitset, max_recid)

        # Excludes - results
        if operator == "-":
            excluded_hits = excluded_hits.union(matched_recs)
        # + and | are interpreted as OR
        else:
            included_hits = included_hits.union(matched_recs)
            all_ranked_results.extend(ranked_result_part)

    ranked_result = []
    if hitset:
        # Removes the excluded records
        result_hits = included_hits.difference(excluded_hits)

        # Avoids duplicate results and normalises scores
        ranked_result = get_greatest_ranked_records(all_ranked_results)
        ranked_result = get_normalised_ranking_scores(ranked_result)
```

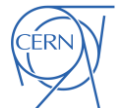

```python
    # Considers not ranked records
    not_ranked = hitset.difference(result_hits)
    if not_ranked:
        lrecIDs = list(not_ranked)
        ranked_result = zip(lrecIDs, [0] * len(lrecIDs))
                            + ranked_result

        ranked_result.sort(lambda x, y: cmp(x[1], y[1]))
        return (ranked_result, params["prefix"], params["postfix"],
                voutput)

    return (ranked_result, "", "", voutput)
```